\documentclass[aps,twocolumn,superscriptaddress,showpacs,floatfix]{revtex4}

\usepackage{graphicx}
\usepackage{dcolumn}
\usepackage{bm}
\usepackage{amsmath2000}

\usepackage{subfigure}

 \newcommand*{\half}{\ensuremath{\frac{1}{2}}}
 \newcommand*{\nd}{\text{ and}}
 \newcommand*{\labe}[1]{\label{e:#1}}
 \newcommand*{\labf}[1]{\label{f:#1}}
 \newcommand*{\labt}[1]{\label{t:#1}}
 \newcommand*{\labs}[1]{\label{s:#1}}
 
 \newcommand*{\eqr}[1]{Eq.~\protect\eqref{e:#1}}
 \newcommand*{\eqs}[1]{Eqs.~\protect\eqref{e:#1}}
 
 \newcommand*{\fig}[1]{Fig.~\protect\ref{f:#1}}
 
 \newcommand*{\tbl}[1]{Table~\protect\ref{t:#1}}
 \newcommand*{\sct}[1]{Section~{\protect\ref{s:#1}}}

\newcommand{\epsf}[2]{
\begin{figure}
\includegraphics[width=\columnwidth]{#1}
\caption{#2\labf{#1}}
\end{figure}}

\newcommand{\epsw}[3]{
\begin{figure*}
\includegraphics[width=#1]{#2}
\caption{#3\labf{#2}}
\end{figure*}}

\newcommand{\epsx}[3]{
\begin{figure}
\includegraphics[width=#1]{#2}
\caption{#3\labf{#2}}
\end{figure}}

\newcommand{\domg}{\ensuremath{\Delta \omega}}
\newcommand{\domq}{\ensuremath{\Delta \omega_q}}
\newcommand{\dndT}{\frac{d\, \eta}{d\, T}}

\DeclareMathAlphabet{\mathsfsl}{OT1}{cmss}{m}{sl}

\begin{document}

\bibliographystyle{apsrev}

\title{Model of Thermal Wavefront Distortion in Interferometric Gravitational-Wave
Detectors I: Thermal Focusing}

\author{R.\ G.\ Beausoleil}
 \altaffiliation[Visiting Scientist; Permanent address: ]{Hewlett-Packard Laboratories,
13837 175$^\textrm{th}$ Pl.\ NE, Redmond, WA 98052--2180}
 \email{beausol@hpl.hp.com}
 \affiliation{Edward L.\ Ginzton
Laboratory, Stanford University, Stanford, CA  94305}
 \author{E.\ D'Ambrosio}
 \affiliation{LIGO Project, California Institute of Technology,
Pasadena, CA  91125}
 \author{W.\ Kells}
 \affiliation{LIGO Project, California Institute of Technology,
Pasadena, CA  91125}
 \author{J.\ Camp}
 \affiliation{LIGO Project, California Institute of Technology,
Pasadena, CA  91125}
 \author{E.\ K.\ Gustafson}
 \affiliation{Edward L.\ Ginzton Laboratory, Stanford University,
Stanford, CA  94305}
 \author{M.\ M.\ Fejer}
 \affiliation{Edward L.\ Ginzton Laboratory, Stanford University,
Stanford, CA  94305}

\date{\today}

\begin{abstract}
We develop a steady-state analytical and numerical model of the
optical response of power-recycled Fabry-Perot Michelson laser
gravitational-wave detectors to thermal focusing in optical
substrates. We assume that the thermal distortions are small
enough that we can represent the unperturbed intracavity field
anywhere in the detector as a linear combination of basis
functions related to the eigenmodes of one of the Fabry-Perot arm
cavities, and we take great care to preserve numerically the
nearly ideal longitudinal phase resonance conditions that would
otherwise be provided by an external servo-locking control system.
We have included the effects of nonlinear thermal focusing due to
power absorption in both the substrates and coatings of the
mirrors and beamsplitter, the effects of a finite mismatch between
the curvatures of the laser wavefront and the mirror surface, and
the diffraction by the mirror aperture at each instance of
reflection and transmission. We demonstrate a detailed numerical
example of this model using the MATLAB program Melody for the
initial LIGO detector in the Hermite-Gauss basis, and compare the
resulting computations of intracavity fields in two special cases
with those of a fast Fourier transform field propagation model.
Additional systematic perturbations (e.g., mirror tilt,
thermoelastic surface deformations, and other optical
imperfections) can be included easily by incorporating the
appropriate operators into the transfer matrices describing
reflection and transmission for the mirrors and beamsplitter.
\end{abstract}


 \maketitle

\section{Introduction}

 In the present decade a number of long-baseline laser
interferometers are expected to become operational, and begin a
search for astrophysical sources of gravitational
radiation.\cite{wil93,bla93,sau94} These include the Laser
Interferometer Gravitational Wave Observatory (LIGO),\cite{abr92,
abr96} the VIRGO project,\cite{gia95} the GEO~600
project,\cite{dan95} and the TAMA~300 project.\cite{tsu95} All of
these detectors will employ a variant of a Michelson
interferometer illuminated with stabilized laser light. The light
will be phase-modulated at one or more radio frequencies,
producing modulation sidebands about the carrier frequency which
provide a phase reference for sensing small variations of the
interferometer arm lengths.\cite{dre83} Gravitational radiation
will produce a differential length change in the arms of the
Michelson interferometer, resulting in a signal at the output
port.

\epsf{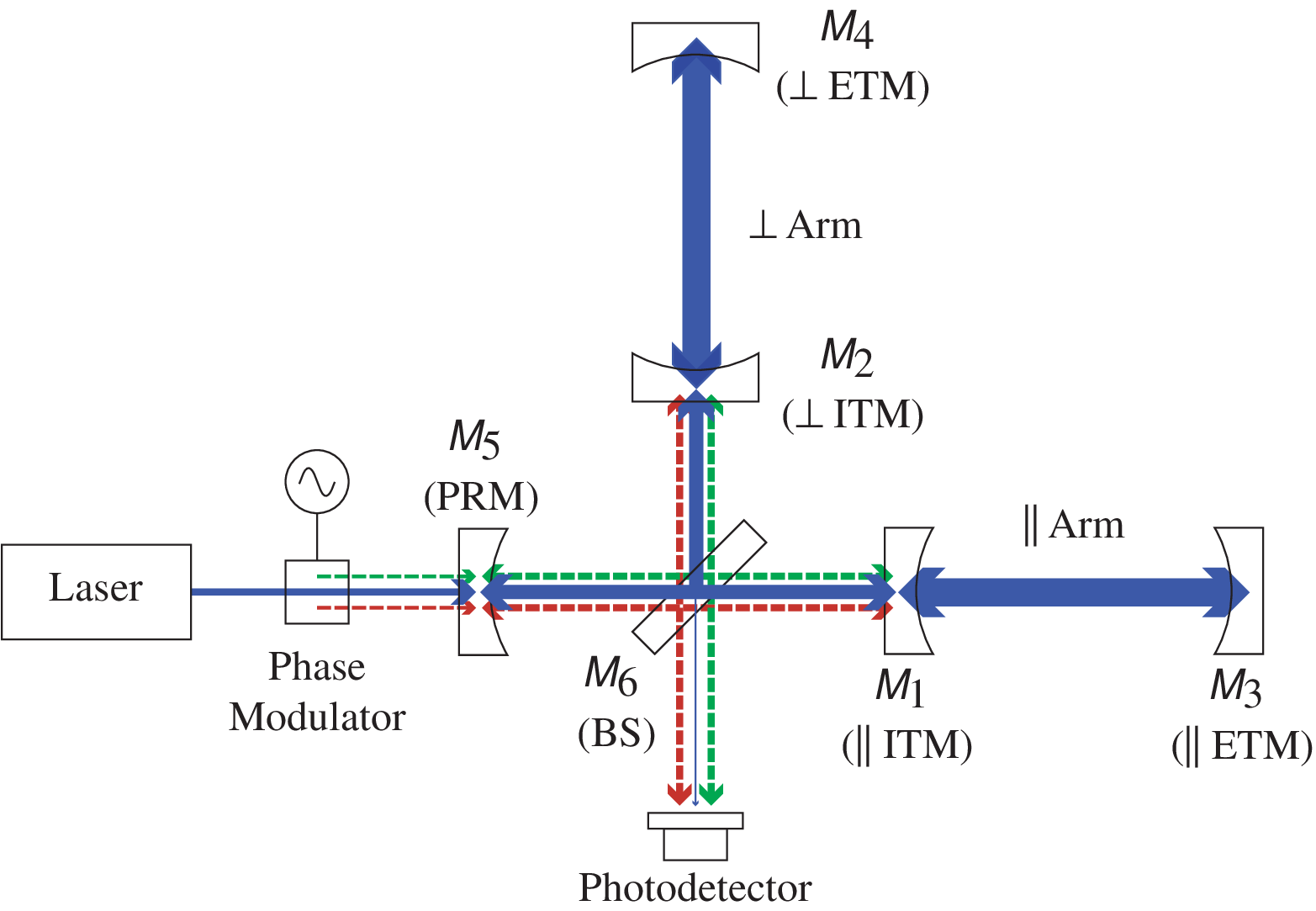}{Schematic configuration of the initial LIGO
detector, a power-recycled Fabry-Perot Michelson interferometer
(PRFPMI).}

In \fig{caltech_prifo}, we show the configuration of the initial
LIGO detector, a power-recycled Fabry-Perot Michelson
interferometer (PRFPMI). The light from a stabilized laser source
enters the interferometer, which is comprised of an asymmetric
Michelson interferometer with Fabry-Perot arm cavities. The arm
cavities consist of polished input and end test masses (ITM and
ETM, respectively) whose coated surfaces also act as mirrors to
resonantly enhance the laser light. An additional power-recycling
mirror (PRM) placed between the laser and beamsplitter increases
the total light power available to the arms, by forming a
``recycling cavity'' together with the beamsplitter and arm cavity
input mirrors. The sideband frequency and interferometer lengths
are adjusted so that the sidebands resonate in the recycling
cavity but not in the arms, while the carrier resonates in both
sets of cavities. This allows extraction of control signals for
all four length degrees of freedom.\cite{fri01}

The noise level in the detection band at the interferometer output
port must be held below the required strain sensitivity. The
primary noise sources limiting the interferometer sensitivity are
seismic noise at frequencies below 100~Hz, thermal noise between
roughly 100 and 300~Hz, and photon counting noise at frequencies
greater than 300~Hz. Photon counting noise is suppressed by using
sufficient laser power; a design strain sensitivity of roughly
10$^{-21}$ over a time period of a millisecond will require a
power of several hundred watts incident on the beamsplitter.

A concern in the operation of an interferometer at such high power
levels is optical distortion. The presence of absorption in the
coatings and substrates of the optics can give rise to non-uniform
heating due to the Gaussian intensity profile of the laser light.
This heating will cause a temperature rise in the substrate,
which, coupled to the temperature dependence of the substrate
index of refraction, will cause a distortion of the wavefront of
the light transmitted through the substrate. For example, the
initial LIGO design provides for fused silica input test masses
which are 10~cm thick. To ensure the stability of the arm cavity
resonant mode, the ITM radius of curvature is set to be several
times the length of the arm cavity, or of order 10~km. By
comparison, the substrate heating due to a nominal substrate and
coating absorption of 5~ppm/cm and 1~ppm respectively causes the
generation of a thermal lens in the substrate with a nominal focal
length of 8~km. This significant distortion can have important
consequences for both the power buildup of the RF sidebands in the
recycling cavity and the interferometer sensitivity.

There have been a number of efforts made to understand the effects
of the propagation of distorted wavefronts through gravitational
wave interferometers, including an approximate treatment of
wavefront distortion in power-recycled and dual-recycled systems
due to imperfections in the shape or alignment of some optical
components,\cite{mee91} an approximate treatment of thermal
lensing and surface deformation in early Fabry-Perot Michelson
designs,\cite{win91} and an approximate description of thermal
lensing in alternate signal extraction designs.\cite{str94,mcc99}
Each of these discussions emphasized the effects of optical
distortion on a particular subset of the optical performance
characteristics of an interferometer. In principle, fast Fourier
transform methods\cite{fft92,fft97,fft98} can be used to simulate
a thermally loaded interferometer, but the computational cost of
an iterated self-consistent FFT numerical analysis of the
corresponding intracavity fields would be prohibitive.

By contrast, we have developed a flexible and robust analytical
and numerical model of the effects of wavefront distortion on the
optical performance of a gravitational-wave interferometer that
requires far fewer computational resources than an FFT approach.
The model includes accurate contributions to the thermal
distortions in all substrates due to optical absorption in the
substrate and both coatings of each test mass,\cite{hel90a} the
effects of any mismatch between the curvatures of the laser
wavefront and the mirror surface at each reflection, and the
diffraction by an aperture at each instance of reflection and
transmission. Operators representing these physical effects on the
reflection and transmission properties of mirrors and
beamsplitters are explicitly incorporated into a set of transfer
(or scattering) matrices that relate output fields to input fields
incident on the surfaces of those elements. The corresponding
transfer matrices of more complex optical components (e.g., a
Fabry-Perot interferometer) can subsequently be constructed in a
hierarchical fashion from those of simpler systems. Using this
approach, the model determines a self-consistent steady-state
solution to the interferometer power buildup in the presence of
optical distortion, and employs an internal numerical mechanism to
preserve the nearly ideal longitudinal phase resonance conditions
that would otherwise be provided by an external servo-locking
control system. We have developed computer code representing this
model that is sufficiently flexible to allow new physical effects
to be added to the simulations with a modicum of effort, and that
can be stopped, interrogated, and restarted at any stage of a
simulation. Our implementation of the artificial resonance-locking
mechanism provides an exponential reduction in the execution time
required for a self-consistent simulation by dynamically adjusting
the frequency response of the interferometer.

Our current model employs basis function expansions of the
electromagnetic field everywhere in the
interferometer.\cite{and84,mor94,hef96,bea99} In principle, these
basis functions can be spatial eigenfunctions of the
interferometer, but in practice, imperfect mode-matching of the
multiple cavities comprising the full interferometer and the
presence of apertures in the system means that round-trip spatial
propagators are \emph{not} Hermitian, and the eigenfunctions are
not power-orthogonal.\cite{sie86,oug87,kos97} Therefore, we choose
an expansion of the intracavity field in an arbitrary set of
power-orthogonal unperturbed spatial basis functions that are not
necessarily eigenfunctions of any subsystem of a thermally loaded,
perturbed interferometer. Since we can move from one spatial
representation to another simply by recomputing the propagator
matrix elements, the relative merit of a set of possible basis
choices is determined by the number of basis modes needed to
describe the intracavity field accurately. Our analysis of the
operators encapsulating the physics of the perturbative phenomena
described in the model is invariant under such a change of spatial
basis. Nevertheless, the choice of an incomplete subset of basis
functions inevitably introduces false optical losses because power
is transferred to higher-order spatial modes which are not
included in a necessarily finite numerical simulation. Therefore,
we introduce unitary approximations for non-diffractive operators
that monotonically improve as more basis functions are included in
the simulation.

In \sct{mocs} of this paper, we describe our representation of the
intracavity electromagnetic field, and the operators incorporated
into the transfer matrices of the subsystems comprising components
of typical interferometric gravitational wave detectors. We
develop a unitary approximation for non-diffractive operators
expanded in finite-dimensional basis sets, and, in our discussions
of the Fabry-Perot and Michelson interferometers, we introduce
abstract idealizations of the control systems needed to maintain
different intracavity resonance conditions. In \sct{prfpmi}, using
these building blocks we construct a model of the initial LIGO
interferometer, and we point out that the basis functions chosen
for a matrix description of the interferometer need not be spatial
eigenfunctions of the fully coupled resonators. As a preliminary
verification of our method, we compare the predictions of our
numerical implementation of this model in a set of well-defined
linear test cases with the output generated by a fast Fourier
transform computer code suite. Finally, we simulate the optical
performance of the initial LIGO interferometer under full thermal
load, using a straightforward nonlinear solution method that
computes the steady-state fields everywhere in the system after
only a few seconds on an ordinary desktop computer.

\section{Models of Optical Components and Structures\labs{mocs}}

\subsection{Representation of the Electromagnetic Field}
Throughout this work, we represent a propagating laser electric
field ${\bf E} ({\bf r}, t)$ with angular carrier frequency
$\omega_0$ as the real part of a product of a real
time-independent polarization unit vector ${\bf e}$, a complex
amplitude function $E({\bf r}, t)$, and the carrier wave function
$\exp\left[ i (k_0 z - \omega_0 t)\right]$, where $\lambda_0
\equiv 2 \pi/k_0 \equiv 2 \pi c/\omega_0$ is the laser
wavelength:\cite{bea99}
 \begin{equation}
{\bf E}({\bf r}, t) \equiv {\rm Re}\left\{ {\bf e}\, E({\bf r}, t)
\exp\left[ i (k_0 z - \omega_0 t)\right] \right\}\labe{Evec}.
 \end{equation}
We use the amplitude function $E({\bf r}, t)$ to represent any
closely-spaced frequency components (such as
radio-frequency-modulated sidebands) as
\begin{equation}
E({\bf r}, t) \equiv \sum_q E_q({\bf r}, t) \exp\left[ i (k_q z -
\omega_q t)\right] ,\labe{Esum}
\end{equation}
where the summation is taken over all frequency components
(including the carrier), $E_q({\bf r}, t)$ is the complex
amplitude function of component $q$, and $\Delta \omega_q \equiv
\omega_q - \omega_0 \equiv \Delta k_q/c \ll \omega_0$ is the
angular frequency shift of component $q$. By convention, the
carrier is labeled by $q = 0$, and therefore $\Delta \omega_0 =
0$. We express all field amplitudes in units of $\sqrt{{\rm
W}/{\rm cm}^2}$, so that the time-averaged intensity carried by
${\bf E} ({\bf r}, t)$ is $\bar{I}({\bf r}, t) = \frac{1}{2}
\sum_q |E_q({\bf r}, t)|^2$.

We will represent the transverse spatial dependence of the field
amplitude $E_q({\bf r}, t)$ at any propagation plane within a
laser interferometric gravitational-wave detector as a linear
superposition of a set of amplitude basis functions with fixed
transverse spatial profiles. In general, these amplitudes are
eigenfunctions of a non-Hermitian integral transform equation
(generally a composition of many consecutive integral transforms,
each of which carries the field from one aperture to the next)
that describes round-trip propagation through a specific resonant
subsystem of the unperturbed interferometer. These resonant
eigenfunctions can then be propagated throughout the remainder of
the interferometer, defining a unique set of spatial basis
functions.\cite{bea99} Therefore, given the existence of a
numerically complete set of these transverse spatial
eigenfunctions, we can expand the sideband amplitude function as
\begin{equation}
E_q({\bf r}, t) = \sum_{m n} E_{m n q}(z, t)\, u_{m n}({\bf r}) .
\labe{emnq}
\end{equation}
Hence, we can convert any composition of two consecutive integral
transforms into a simple matrix product. In practice, we are
interested in the steady-state characteristics of the detector
under thermal load, and we will clearly indicate the propagation
plane under discussion, so we will usually suppress the explicit
coordinates $\{z, t\}$ in later sections.

As an example, consider the simple case of laser field propagation
through a distance $L \equiv z_2 - z_1$ in vacuum. Suppose that we
choose to represent the spatial and temporal properties of this
field using $N$ transverse eigenfunctions $u_{m n}({\bf r})$ and
$Q$ frequency components, including the carrier and both positive
and negative sidebands for each nonzero radio modulation
frequency. We arrange the expansion coefficients into an $N \times
Q$ ordered matrix $E$ with elements $E_{m n q}$, where each row
represents the spatial eigenfunction corresponding to a particular
choice of $\{m, n\}$ and each column corresponds to a particular
frequency component $q$, and we assume that $L$ has been chosen to
provide resonant excitation of the fundamental carrier spatial
mode $E_{000}$. In this basis, the spatial contribution to the
vacuum propagation kernel is given by a $N \times N$ diagonal
matrix $G$ with elements
\begin{equation}
G_{m n, m^\prime n^\prime} = \exp(i\, \Delta\, \varphi_{m n})\,
\delta_{m m^\prime}\, \delta_{n n^\prime} ,
\end{equation}
where $\Delta\, \varphi_{m n} \equiv \varphi_{m n} - \varphi_{0
0}$ is the net Gouy phase (relative to the fundamental spatial
mode) accumulated by mode $m n$ over the propagation distance $L$,
and the net longitudinal optical path length contribution
(relative to the carrier) is given by a $Q \times Q$ diagonal
matrix $\Omega$ with elements
\begin{equation}
\Omega_{q q^\prime} = \exp(i \Delta \omega_q \tau)\, \delta_{q
q^\prime} ,
\end{equation}
where $\tau \equiv L/c$ is the propagation time. The components of
the field following the propagation step can then be computed
using the sparse matrix product
\begin{equation} \labe{freespace}
E(z_2) = G\, E(z_1)\, \Omega .
\end{equation}

We are particularly interested in the more complex example of a
power-recycled Fabry-Perot Michelson laser gravitational-wave
interferometer. In its unperturbed state (i.e., in the absence of
aperture diffraction and thermal wavefront distortions), this
detector configuration is described well by Hermite-Gauss basis
functions under perfect mode-matching conditions. The
infinite-aperture Hermite-Gauss basis is orthonormal (since it is
generated by a Hermitian propagation equation), so that field
amplitude expansions using this basis are power
orthogonal,\cite{sie86} allowing time-averaged intensities to be
calculated using the sum of the inner products $|E_q({\bf r},
t)|^2 = \sum_{m\, n}\, |E_{m n q}(z, t)|^2$. If both Gouy and
optical path length phase contributions are accumulated using
\eqr{freespace}, then we can represent an arbitrary non-astigmatic
Hermite-Gauss basis function $u_{m n}(x, y)$ at a given reference
plane using
 \begin{equation} \labe{umndef}
 \begin{split}
 u_{m n}(x, y) &= \left( \frac{1}{2^{m + n} m! n!}\, \frac{2}{\pi\, w^2}
 \right)^{1/2} \\
 &\times H_m\left(\frac{\sqrt{2} x}{w}\right)\,  H_n\left(\frac{\sqrt{2}
 y}{w}\right)\, \\
 &\times \exp\left[ \left(i \frac{k}{2 R} - \frac{1}{w^2}\right) \left(x^2 +
 y^2\right)\right],
 \end{split}
 \end{equation}
 where $H_n(x)$ is the Hermite polynomial of order $n$, $w$ is the beam spot radius,
 and $R$ is the wavefront radius of
 curvature at that reference plane. Throughout this work, we will
 define the $x$-$z$ plane to be ``horizontal,'' containing the central beam
 paths of the two Fabry-Perot arm cavities in \fig{caltech_prifo},
 and the $y$-$z$ plane to be ``vertical,'' perpendicular to the
 horizontal plane and parallel to the local gravity gradient of
 the Earth.

\subsection{Mirrors\labs{mirrors}}
\subsubsection{Aperture Diffraction and Mirror-Field Curvature Mismatch Operators\labs{mir_aper_curv}}
Consider the general problem of the reflection of a propagating
electromagnetic field by a perfectly reflecting cylindrically
symmetric mirror $\mathcal{M}$ with spherical radius of curvature
$R_M$ and aperture diameter $2 a$. The reflection can be described
in a single step taken from the aperture $\mathcal{A}$ at the
reference plane $z^<$ conventionally located just \emph{before}
the reflection from $\mathcal{M}$ to the field position $z^>$ just
\emph{after} reflection has occurred. If the transverse field at
$z^<$ is $E(x^\prime, y^\prime, z^<)$, then the corresponding
field at $z^>$ is given by Huygens's integral in the Fresnel
approximation as\cite{oug87}
\begin{equation} \labe{genprop}
E(x, y, z^>) = \int_{\mathcal{A}} dx^\prime\, dy^\prime\, K(x, y;
x^\prime, y^\prime) E(x^\prime, y^\prime, z^<) ,
\end{equation}
where the functional form of the forward propagation kernel for a
cylindrically symmetric mirror is\cite{bea99}
\begin{equation} \labe{KRM}
\begin{split}
 K(x, y; x^\prime, y^\prime) &= -
\exp\left[-i\,\frac{\pi}{\lambda_0} \frac{2}{R_M}
\left({x^\prime}^2 + {y^\prime}^2\right)\right] \\ &\times\,
\delta(x - x^\prime)\, \delta(y - y^\prime) .
\end{split}
\end{equation}
Here the leading negative sign accounts for the Maxwell boundary
condition at the perfectly reflecting mirror surface.

If we expand the field at both reference planes using a
numerically complete set of cartesian basis functions as in
\eqr{emnq}, then we can compute the matrix elements of the forward
propagation kernel in that basis as
\begin{equation}\labe{genca}
\begin{split}
K_{m n, m^\prime n^\prime} &=
\iint\limits^{\quad\;\infty}_{-\infty} dx\,dy\,
\iint_{\mathcal{A}} dx^\prime\,dy^\prime\; K(x, y; x^\prime,
y^\prime) \\ &\qquad\times\, u^\dagger_{m n}(x, y, z^> )\,
u_{m^\prime n^\prime}(x^\prime, y^\prime, z^<)\\
 &= \iint_{\mathcal{A}} dx\,dy\; \exp\left[-i\,\frac{\pi}{\lambda_0} \frac{2}{R_M}
\left(x^2 + y^2\right)\right] \\ &\qquad\times\, u^\dagger_{m
n}(x, y, z^>)\, u_{m^\prime n^\prime}(x, y, z^<)
\end{split}
\end{equation}
Here the function $u^\dagger_{m n}(x, y, z^>)$ is \emph{not}
generally the conjugate transpose of $u_{m n}(x, y, z^>)$.
Instead, these basis functions represent fields propagating
through the system in the \emph{reverse} direction, and are
obtained by solving the Huygens-Fresnel integral \eqr{genprop}
using $K^T(x, y; x^\prime, y^\prime)$, the transpose of the
forward-propagation kernel.\cite{bea99, oug87, sie86}

Although the reflection-diffraction problem is specified
completely by \eqr{genca}, it can introduce inconsistencies in
practical implementations of optical resonator simulations. For
example, when a finite number of basis functions are used to
represent the propagating field, the matrix calculated using
\eqr{genca} generates two sources of optical loss: genuine
diffraction loss due to field truncation by the aperture, and
power transferred to higher-order optical modes which are ignored
in the simulation. The latter contribution is unphysical: in the
limit $a \rightarrow \infty$, when a numerically complete set of
basis functions is used the propagation matrix must conserve
energy (i.e., must be unitary.) Therefore, for the purposes of our
simulations we will explicitly allow diffraction loss due to
aperture truncation only, and separate the effects of aperture
diffraction and reflection into two consecutive propagation steps
using the formulation
\begin{equation} \labe{kca}
K \cong C\, A ,
\end{equation}
where $A$ is a matrix operator representing purely transmissive
aperture diffraction, and $C$ is a unitary matrix operator
describing direct focusing by a spherical mirror at normal
incidence.

As a concrete example, we first calculate the matrix elements of
$A$ in the Hermite-Gauss basis, treating the truncation as a
perturbation and describing the field both before and after the
aperture using a linear superposition of a finite set of
unperturbed Hermite-Gauss functions. Therefore, $u^\dagger_{m
n}(x, y) = u^*_{m n}(x, y)$, and we define the analytical integral
\begin{equation} \labe{Adef}
\begin{split}
A_{m n, m^\prime n^\prime} &= \iint_{\mathcal{A}} dx\,dy\; u^*_{m
n}(x, y)\, u_{m^\prime n^\prime}(x, y)\\
 &\equiv \delta_{m, m^\prime}\, \delta_{n, n^\prime} - e^{-\alpha}\, I_{m n, m^\prime
 n^\prime}(\alpha),
\end{split}
\end{equation}
where $\alpha \equiv 2(a/w)^2$, and the integration is performed
over the circular aperture defined by $\sqrt{x^2 + y^2} \le a$. We
specify a basis set by designating the transverse modes for a
given frequency component $q$ in order of increasing $m + n$
first, and then in order of increasing $n$. For example, if we use
a Hermite-Gauss basis with $m + n \le 2$, then for the column
vector $E_q \equiv \{E_{00q}, E_{10q}, E_{01q}, E_{20q}, E_{11q},
E_{02q}\}^T$ we obtain the matrix
\begin{widetext}
\begin{equation}
I(\alpha) = \begin{bmatrix}
    1 & 0 & 0 & \frac{1}{\sqrt{2}} \alpha & 0 & \frac{1}{\sqrt{2}} \alpha \\
    0 & 1 + \alpha & 0 & 0 & 0 & 0 \\
    0 & 0 & 1 + \alpha & 0 & 0 & 0 \\
    \frac{1}{\sqrt{2}} \alpha & 0 & 0 & 1 + \half \alpha + \frac{3}{4}\alpha^2 & 0 & \frac{1}{4} \alpha (\alpha - 2) \\
    0 & 0 & 0 & 0 & 1 + \alpha + \half\alpha^2& 0 \\
    \frac{1}{\sqrt{2}} \alpha & 0 & 0 & \frac{1}{4} \alpha (\alpha - 2) & 0 & 1 + \half \alpha + \frac{3}{4}\alpha^2
    \end{bmatrix} ,
\end{equation}
 \end{widetext}
Note that nonzero elements of $I(\alpha)$ satisfy the relation
$I_{m n, m^\prime n^\prime}(\alpha) \approx O[\alpha^{\half (m + n
+ m^\prime + n^\prime)}]$.

Next, we ignore aperture diffraction and calculate the matrix
elements of $C$ in the Hermite-Gauss basis. Strictly speaking, the
matrix elements of $C$ can be calculated using \eqr{genca} in the
limit $a \rightarrow \infty$. However, as described above, we seek
an explicitly unitary approximation for $C$ that monotonically
improves as we include more basis functions in the field
expansion. When \eqr{genca} is applied to a Hermite-Gauss basis
function with field curvature parameter $R_F$, this parameter is
transformed according to the $ABCD$ propagation law as
$1/R_F^\prime = 1/R_F - 2/R_M$. In the unperturbed case where $R_M
= R_F$, the curvature parameter of the backward-propagating basis
function is therefore $-(-R_F) = R_F$, and using \eqr{umndef} we
have
\begin{equation} \labe{Cdef}
C \cong \exp(i\, \gamma\, c) ,
\end{equation}
where
\begin{equation} \labe{gammadef}
\gamma \equiv \frac{\pi\, w^2}{\lambda} \left(\frac{1}{R_F} -
\frac{1}{R_M}\right) ,
\end{equation}
and
\begin{equation}
\begin{split}
c_{m n, m^\prime n^\prime} &\equiv \frac{2}{w^2}
\iint\limits^{\quad\;\infty}_{-\infty} dx\,dy\;  \, \left(x^2 +
y^2\right)\\ &\times\, \left| u_{m n}(x, y)\, u_{m^\prime
n^\prime}(x, y)\right| .
\end{split}
\end{equation}
As the number of Hermite-Gauss basis functions used to represent
$E_q$ grows, $\exp(i\, \gamma\, c) \rightarrow C$, and since $c$
is symmetric, $\exp(i\, \gamma\, c)$ is explicitly unitary. Using
the generating function of the Hermite polynomials, we obtain the
matrix elements
\begin{equation}
c_{m n, m^\prime n^\prime} = X^2_{m, m^\prime}\, \delta_{n,
n^\prime} + \delta_{m, m^\prime}\, X^2_{n, n^\prime} ,
\end{equation}
where
\begin{equation}
\begin{split}
X^2_{m, m^\prime} &\equiv \half (1 + 2 m)\, \delta_{m, m^\prime}\\
&+ \half \sqrt{(m^\prime + 1)(m^\prime + 2)}\; \delta_{m, m^\prime
+ 2} \\ &+ \half \sqrt{(m + 1)(m + 2)}\; \delta_{m^\prime, m + 2}.
\end{split}
\end{equation}

As a straightforward test of our unitary approximation, we can
compare the separated propagator given by \eqr{kca} with the exact
propagator specified by \eqr{genca} for the TEM$_{00}$
Hermite-Gauss mode. An explicit calculation using \eqr{genca}
yields
\begin{equation} \labe{K0000exact}
K_{00,00} = \frac{1 - e^{-(1 + i \gamma) \alpha}}{1 + i \gamma} .
\end{equation}
In \fig{ca_error}, we plot the fractional error in $| K_{00,00} |$
accrued by computing $K = C\, A$, where $A$ is provided by
\eqr{Adef} and $C$ by \eqr{Cdef}. For an incident field with a
spot radius $w = 3.64$~cm and a wavefront radius of curvature $R_F
= 10$~km at a wavelength $\lambda = 1.0642$~$\mu$m, we show
relative errors for different values of the mirror's cylindrical
substrate radius and radius of curvature as a function of the
maximum mode index $N$. The number of basis modes with $m + n \le
N$ used in the calculations of $A$ and $C$ is given by $\half (N +
1) (N + 2)$. As we expect, the relative error decreases
significantly as the number of basis modes increases.

\epsf{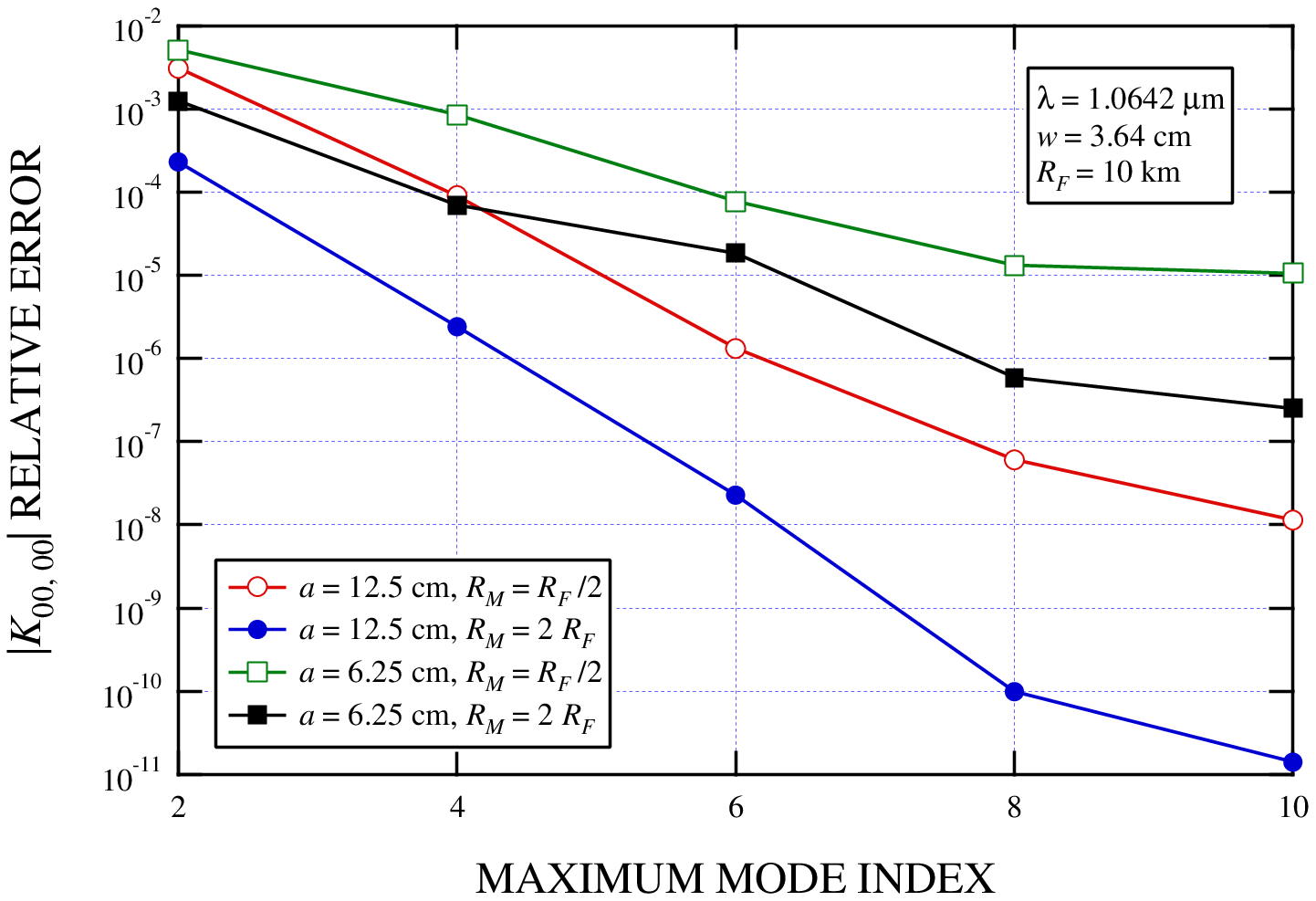}{The fractional error in $| K_{00,00} |$ obtained
by comparing the exact value given by \eqr{K0000exact} with the
TEM$_{00}$ element of the approximate propagator $K = C\, A$,
where $A$ is provided by \eqr{Adef} and $C$ by \eqr{Cdef}.}

 \subsubsection{Thermal Focusing Operators\labs{mir_thermal_lens}}
In an interferometric gravitational-wave detector, the test masses
(as well as any optics which form power and/or signal recycling
cavities) are thick cylindrical blocks of a transparent material,
such as pure fused silica. A high-reflectance coating (HR) is
applied to the broad inner face of the cylinder, forming a mirror;
if transmission through the mirror substrate is necessary, then an
antireflecting coating (AR) will be applied to the outer surface
as well. Under test, then, laser power may be absorbed in the
coatings and/or the mirror substrate (SS), and the corresponding
heat flow into the substrate will result in an inhomogeneous
temperature distribution throughout the mirror. This temperature
gradient will cause a refractive index gradient to form,
converting the substrate into a thermally-generated thick lens.

In general, the propagation of the laser field through the
substrate will be described by an integral transform equation that
incorporates the effects of position-dependent optical phase
shifts caused by the thermal load. As discussed above, by choosing
a suitable set of unperturbed basis functions (i.e., in the
absence of thermal and mechanical perturbations) we can convert
the integral transform equation into a scattering matrix operator
which redistributes energy contained in an initial configuration
of basis functions into the final set of corresponding functions
that emerges from the substrate. In order to determine the
elements of this matrix operator, we must first find the
temperature distribution in the substrate due to absorption in
both the coatings and the substrate, and then we must compute the
matrix elements of the longitudinal optical path-difference (OPD)
phase shift introduced by the temperature gradient.

Hello and Vinet\cite{hel90a} have calculated both the steady-state
and transient temperature distribution throughout the simplified
mirror shown in \fig{mirror_hv} for the case where the change in
the temperature due to the absorption of optical power remains
small compared to the external ambient temperature $T_0$. This
assumption linearizes the radiative boundary conditions at the
surfaces of the mirror, and allows the temperature increase due to
coating and substrate absorption to be calculated independently
(and then summed) for some incident power $P$. As shown in
\fig{mirror_hv}, they treated the mirror substrate as a thick disk
spanning the cylindrical coordinates $0 \le r \le a$ and $-h/2 \le
z \le h/2$, and the coating is concentrated in an infinitesimally
thin layer at $z = -h/2$. After an algebraic simplification of
Hello and Vinet's steady-state results, we find that the
temperature distributions throughout the substrate due to
absorption in the substrate and coating are, respectively,
\begin{equation} \labe{Ts}
\begin{split}
T_s(r, z) &= P\, \frac{\alpha_s a^2}{k_T} \sum_k
\frac{p_k}{\zeta_k^2} \\
&\times \biggl[1 - 2 \tau A_k \cosh\left(\zeta_k\, \frac{z}{a}\biggr)%
\right] J_0\left(\zeta_k\, \frac{r}{a}\right) ,
\end{split}
\end{equation}
\begin{equation} \labe{Tc}
\begin{split}
&T_c(r, z) = P\, \frac{a_c a}{k_T} \sum_k p_k \\ &\times\,
\biggl[A_k \cosh\left(\zeta_k\, \frac{z}{a}\right) - B_k
\sinh\left(\zeta_k\, \frac{z}{a}\right)\biggr] J_0\left(\zeta_k\,
\frac{r}{a}\right) ,
\end{split}
\end{equation}
where $a_c$ is the power absorption of the coating, $\alpha_s$ is
the absorption coefficient of the substrate material,
\begin{subequations}
\begin{eqnarray}
A_k &=& \frac{1}{2\left[\zeta_k \sinh\left(\gamma_k\right) + \tau
\cosh\left(\gamma_k\right)\right]} , \\ B_k &=&
\frac{1}{2\left[\zeta_k \cosh\left(\gamma_k\right) + \tau
\sinh\left(\gamma_k\right)\right]} ,
\end{eqnarray}
\end{subequations}
$\gamma_k \equiv \zeta_k h/2 a$, $\tau \equiv 4 \epsilon \sigma
T_0^3 a/k_T$, $\sigma$ is the Stefan-Boltzmann constant,
$\epsilon$ is the total spherical emissivity of the substrate, and
$k_T$ is the thermal conductivity of the substrate. For $a =
12.5$~cm (consistent with the value chosen for the LIGO test
masses) and $T_0 = 300$~K, we have $\tau = 0.277$. The terms in
these series are characterized by the roots of the equation
\begin{equation} \labe{zetadef}
\zeta J_1\left(\zeta\right) - \tau J_0\left(\zeta\right) = 0 ,
\end{equation}
which are reasonably well-approximated by the zeros of
$J_1\left(\zeta\right)$, $\zeta_k \cong (k + 1/4)\pi$ for $k \in
\{0, 1, 2, \ldots\}$, when $\tau < 1$.

\epsf{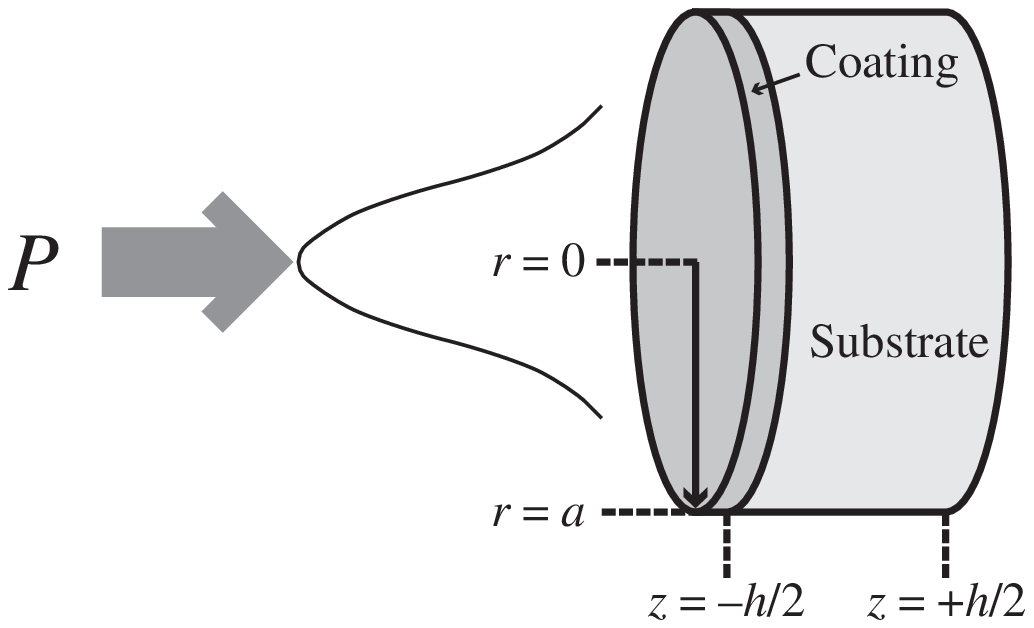}{Schematic diagram of a singly-coated mirror and
the corresponding coordinate system used by Hello and
Vinet\cite{hel90a}. Both the mirror and the incident power
distribution are assumed to be azimuthally symmetric. The total
incident power $P$ can be absorbed in the coating and/or the
substrate, resulting in a total temperature distribution $T(r, z)$
and a corresponding thermal lens in the substrate.}

The constants $p_k$ are the coefficients of a Dini series
expansion\cite{hel90a} of the azimuthally symmetric incident
intensity distribution $I(r)$, given by
\begin{equation} \labe{dini}
I(r) \equiv \sum_k p_k\, J_0\left(\zeta_k\, \frac{r}{a}\right).
\end{equation}
A straightforward calculation using \eqr{zetadef} yields
\begin{equation} \labe{pkdef}
p_k = \frac{2 \zeta_k^2}{\left(\zeta_k^2 + \tau^2\right)
J_0^2\left(\zeta_k\right)} \frac{1}{a^2} \int_0^a dr\, r\,
J_0\left(\zeta_k\, \frac{r}{a}\right) I(r) .
\end{equation}
If we assume that the field is primarily comprised of the
fundamental $\{00\}$ mode, then we can neglect the slight heating
arising from higher-order modes. When the incident intensity
describes a unit power TEM$_{00}$ Gaussian beam with spot radius
$w$, we have $I(r) = (2/\pi w^2) \exp\left(-2 r^2/w^2\right)$.
Furthermore, if $w/a \ll 1$, we can extend the upper limit of
integration in \eqr{pkdef} to infinity, thereby obtaining the
analytic expression
\begin{equation} \labe{pkg}
p_k = \frac{1}{\pi a^2} \frac{\zeta_k^2}{\left(\zeta_k^2 +
\tau^2\right) J_0^2\left(\zeta_k\right)} \exp(-\beta_k) ,
\end{equation}
where $\beta_k \equiv \frac{1}{8} (\zeta_k w/a)^2$.

The $z$ dependence of the substrate temperature distribution given
by \eqr{Ts} is typically very weak when $w/a \ll 1$ and the
boundary conditions are radiative.\cite{hel90a} In this case, near
$r = 0$ we can use the identity $\sum_k p_k = I(0) = 2/\pi w^2$ to
obtain the approximate temperature distribution $T_s(r) - T_s(0)
\cong (\alpha_p P_s/2 \pi k_T) (r/w)^2$. In the thin lens
approximation, this quadratic temperature gradient yields a weak
focal length\cite{sie86}
\begin{equation}
f = \frac{\pi w^2}{\alpha_s h P} \, \frac{k_T}{d \eta/dT} .
\end{equation}
The material constants for fused silica are shown in \tbl{silica}.
If we assume typical LIGO parameter values for the test masses,
then we have $w = 4$~cm, $h = 10$~cm, and $P = 500$~W, giving $f =
16$~km, a length only four times greater than that of a LIGO
Fabry-Perot arm cavity.

\begin{table}
\caption{Material constants for fused silica used in our
simulations.\labt{silica}}
\begin{ruledtabular}
\begin{tabular}{clcc}
Constant & \multicolumn{1}{c}{Name} & Value & Units\\ \hline
 $\eta$     & Refractive index & 1.44968 & \\
 $d \eta/d T$ & First-order $\eta$-$T$ coefficient& $8.7\, \times\, 10^{-6}$ & K$^{-1}$ \\
 $k_T$ & Conductivity & 1.38 & W/m~K \\
 $\epsilon$ & Total spherical emissivity & 0.76 & \\
 $\alpha_s$ & Bulk absorption coefficient & $5\, \times\, 10^{-4}$ &
 m$^{-1}$ \\
 $\sigma_s$ & Bulk scattering coefficient & $0.5\, \times\, 10^{-4}$ &
 m$^{-1}$ \\
\end{tabular}
\end{ruledtabular}
\end{table}

The thermal load on the substrate causes a position-dependent
phase shift across any wavefront propagating through the mirror.
Defining the longitudinal optical path-difference (OPD) phase
$\phi$ as
\begin{equation}
\phi(r) \equiv \frac{2 \pi}{\lambda_0} \dndT \int_{-h/2}^{h/2} d
z\ T(r, z) ,
\end{equation}
we obtain for the substrate and coating contributions
\begin{subequations}
\begin{eqnarray} \labe{OPDs}
\phi_s(r) &=& P_s\, \frac{2 \pi a^2}{\lambda_0 k_T} \dndT \sum_k
\frac{p_k}{\zeta_k^2} \nonumber \\ &\times& \left[1 - \frac{2 \tau
A_k}{\gamma_k} \sinh\left(\gamma_k\right)\right]
J_0\left(\zeta_k\, \frac{r}{a}\right) ,\\ \phi_c(r) &=& P_c\,
\frac{2 \pi a^2}{\lambda_0 k_T} \dndT \sum_k \frac{p_k}{\zeta_k}
\nonumber \\ &\times& 2 A_k \sinh\left(\gamma_k\right)
J_0\left(\zeta_k\, \frac{r}{a}\right) ,
\end{eqnarray}
\end{subequations}
where $P_s \equiv \alpha_p h P$ and $P_c \equiv a_c P$ are the
powers absorbed in the substrate and coating, respectively. If the
thickness of the substrate is small compared to the wavefront
radius of curvature, then we can approximate the propagator
describing transmission through the substrate as a diffraction
step through the aperture followed by a direct integration of the
OPD across the aperture to describe the accumulated distortion. If
we define the ``edge-to-center'' OPD phase as $\Delta \phi(r)
\equiv \phi(r) - \phi(0)$, then we can compute the total substrate
thermal distortion matrix operator using
\begin{equation}
\begin{split}
S_{m n, m^\prime n^\prime} &=
\iint\limits^{\quad\;\infty}_{-\infty} dx\,dy\;  \exp\left[i
\Delta \phi(r) \right] \\ &\times\, u_{m n}^\dagger(x, y)\,
u_{m^\prime n^\prime}(x, y)
\end{split}
\end{equation}
for both substrate and coating absorption.

\epsx{2.75in}{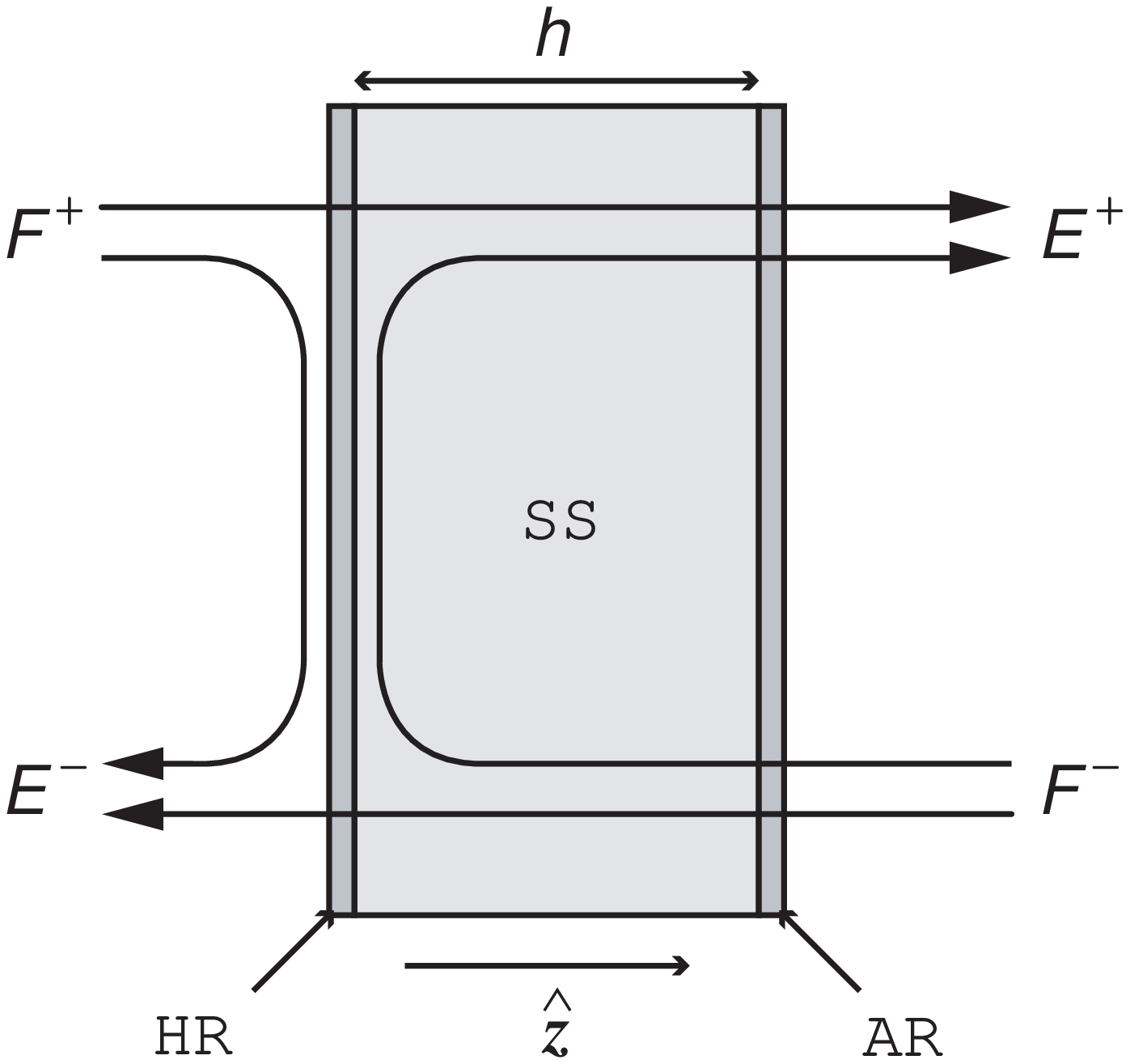}{Schematic diagram of field distributions
in a mirror of thickness $h$, consisting of a substrate (SS) with
applied high-reflectance (HR) and antireflecting (AR) coatings.}

As we discussed above for the wavefront-mirror curvature matrix
$C$ given by \eqr{Cdef}, we seek a unitary approximation of the
substrate thermal distortion matrix to ensure that power is
conserved even when a finite set of unperturbed basis functions is
used to compute the matrix elements. Once again, we begin by
computing the matrix elements of the exponential arguments $\Delta
\phi_s(r)$ and $\Delta \phi_c(r)$ as
\begin{subequations}
 \labe{Phidef}
\begin{eqnarray}
 \labe{Phis} \Phi_{s; m n, m^\prime n^\prime} &=& P_s\, \frac{2
\pi a^2}{\lambda_0 k_T} \dndT \sum_k \frac{p_k}{\zeta_k^2}
\biggl[1 \nonumber \\  &-& \frac{\tau}{\gamma_k}\, 2 A_k
\sinh\left(\gamma_k\right)\biggr] e^{-\beta_k} I_{k; m n, m^\prime
n^\prime}
\\
 \labe{Phic} \Phi_{c; m n, m^\prime n^\prime} &=& P_c\, \frac{2
\pi a^2}{\lambda_0 k_T} \dndT \sum_k \frac{p_k}{\zeta_k}\,
\nonumber \\ &\times& 2 A_k \sinh\left(\gamma_k\right)
e^{-\beta_k} I_{k; m n, m^\prime n^\prime} .
\end{eqnarray}
\end{subequations}
where we have defined the matrix integral $I_k$ as
\begin{equation}\labe{I_kmn}
\begin{split}
I_{k; m n, m^\prime n^\prime} &\equiv e^{\beta_k}
\iint\limits^{\quad\;\infty}_{-\infty} dx\,dy\;
\left[J_0\left(\zeta_k\, \frac{r}{a}\right) - 1\right] \\
&\times\, u_{m^\prime n^\prime}^\dagger(x, y)\, u_{m n}(x, y).
\end{split}
\end{equation}
Since the integrand is zero at $r = \sqrt{x^2 + y^2} = 0$, our
definition of $\Delta \phi(r)$ allows us to explicitly ignore any
phase shift accumulated due to the peak thermal change in the
substrate refractive index, and instead we include only the
contribution to the wavefront distortion arising from the
\emph{shape} of the temperature distribution. Using the same
example Hermite-Gauss basis described in \sct{mir_aper_curv}, for
$m + n \le 2$ and a given value of $k$, we obtain the square
matrix elements
\begin{widetext}
\begin{equation}
I_k(\beta_k) = \begin{bmatrix}
    1 & 0 & 0 & -\,\frac{1}{\sqrt{2}}\,\beta_k & 0 & -\,\frac{1}{\sqrt{2}}\,\beta_k \\
    0 & 1 - \beta_k & 0 & 0 & 0 & 0 \\
    0 & 0 & 1 - \beta_k & 0 & 0 & 0 \\
    -\,\frac{1}{\sqrt{2}}\,\beta_k & 0 & 0 & 1 - 2 \beta_k + \frac{3}{4}\beta_k^2 & 0 & \frac{1}{4} \beta_k^2 \\
    0 & 0 & 0 & 0 & 1 - 2 \beta_k + \frac{1}{2} \beta_k^2& 0 \\
    -\,\frac{1}{\sqrt{2}}\,\beta_k & 0 & 0 & \frac{1}{4} \beta_k^2 & 0 & 1 - 2 \beta_k + \frac{3}{4}\beta_k^2
    \end{bmatrix} ,
\end{equation}
\end{widetext}
Note that nonzero elements of $I_k(\beta_k)$ satisfy the relation
$I_{k; m n, m^\prime n^\prime}(\beta_k) \approx O[\beta_k^{\half
(m + n + m^\prime + n^\prime)}]$.

Collecting results, we can compute the net effective substrate OPD
matrix operator in our unitary approximation as
\begin{equation} \labe{S}
 S = \exp \left\{i\, \left[P_s\, \Phi_s  + \left(P_h +
 P_a\right) \Phi_c \right]\right\} ,
\end{equation}
where $P_s$, $P_h$, and $P_a$ are the total TEM$_{00}$ optical
powers absorbed in the substrate, HR coating, and AR coating,
respectively. The power distribution throughout the mirror volume
is shown schematically in \fig{mirror_abs} for a dual-coated
cylindrical mirror with substrate thickness $h$. A laser field
$F^+$ is incident on the HR coating (the ``front'' of the mirror),
and another field $F^-$ is incident on the AR coating (the
``back'' of the mirror). Since we have assumed that the field is
primarily comprised of the fundamental $\{00\}$ mode, then the
total power absorbed in the coatings and substrate depend only on
the field amplitude expansion coefficients $F_{0 0 q}$, as listed
in \tbl{mirabs}. Here $\ensuremath{a_\mathit{hr}}$ and
$\ensuremath{a_\mathit{ar}}$ are the dimensionless fractional
power absorptions of the HR and AR, respectively, and $\alpha_s$
is the power absorption coefficient of the substrate, with units
of inverse length. The reflected field $E^-$ is the superposition
of the prompt reflection of $F^+$ and the residual of $F^-$ which
has been transmitted through the HR.

\begin{table}
\caption{Total absorbed TEM$_{00}$ powers for each of the three
regions shown in \fig{mirror_abs}, summed over the lowest-order
transverse modes of all active sidebands. The factors of $\half$
arise from our normalization convention for the laser electric
field.\labt{mirabs}}
\begin{ruledtabular}
\begin{tabular}{cl}
Region & \multicolumn{1}{c}{Absorbed Power} \\ \hline
 SS     & $P_s = \alpha_s h \, \half \sum_q \left(|E^+_{00q}|^2 + |F^-_{00q}|^2\right)$
\\
 HR     & $P_h = \ensuremath{a_\mathit{hr}}\, \half \sum_q \left(|F^+_{00q}|^2 +
|F^-_{00q}|^2\right)$ \\
 AR     & $P_a = \ensuremath{a_\mathit{ar}}\, \half \sum_q \left(|E^+_{00q}|^2 +
|F^-_{00q}|^2\right)$ \\
\end{tabular}
\end{ruledtabular}
\end{table}


\subsubsection{Transfer Matrix\labs{mxfer}}
We can extend our analysis of the reflection and transmission
processes and build a general transfer matrix for fields incident
on the mirror from either direction. First, we choose the
laser-industry standard conventions for the phases of the
amplitude reflection and transmission coefficients for a
quarter-wave dielectric stack. Therefore, the Maxwell boundary
conditions for the HR generate the amplitude coefficients $-r$ for
reflection and $i\, t$ for transmission, regardless of propagation
direction. Losses in the HR due to absorption and scattering are
included implicitly when $r^2 + t^2 < 1$. However, we must
explicitly account for losses due to transmission through the
substrate and AR coating using the amplitude coefficient
 \begin{equation}
 \labe{mirsubtrans}
 t_s = e^{-(\alpha_s + \sigma_s)h/2} \sqrt{1 - (a_{ar} + s_{ar})} ,
 \end{equation}
 where $\sigma_s$ is the scattering loss per unit length in the
 substrate, and $s_{ar}$ is the scattering loss in the AR coating.

The propagator for reflection from the vacuum side of the HR is
given by \eqr{KRM}, and the resulting perturbation matrix for a
mismatch between the radius of curvature of the mirror and the
wavefront is given by \eqr{Cdef} as $C(\gamma) \equiv C^-$. Since
the propagator for reflection from the substrate side of the HR is
also given by \eqr{KRM} with $-2/R_M \rightarrow 2
n/R_M$,\cite{sie86} where $n$ is the refractive index of the
substrate, the corresponding perturbation matrix is $C(-n\,
\gamma) \equiv C^+$. Similarly, the propagator for transmission
through the HR from the substrate to the vacuum is given by
\eqr{Cdef} with $-2/R_M \rightarrow (n - 1)/R_M$,\cite{sie86} so
the transmission curvature mismatch matrix operator is $C[(n - 1)
\gamma/2] = (C^- C^+)^{1/2}$.

In \sct{FPIpseudolocker} and \sct{IFOpseudolocker}, we will adjust
the microscopic positions of some of the mirrors to ensure that
the perturbed interferometers maintain the appropriate resonant
phase conditions as the system evolves. If the mirror in
\fig{mirror_abs} is moved in the positive $\hat{z}$ direction
(i.e., toward the right in the diagram) by the microdisplacement
$\Delta z \lesssim \lambda$, then reflection from the ``front'' of
the mirror introduces a relative phase $\exp(+i 2 k \Delta z)$,
while reflection from the ``back'' of the mirror requires a phase
adjustment of $\exp(-i 2 k \Delta z)$.

Collecting these results, the final mirror transfer matrix
depicted in \fig{mirror_xfer} is
 \begin{equation} \labe{mxfer}
 \begin{bmatrix} E^- \\ E^+ \end{bmatrix} =
 \begin{bmatrix} T^- & R^-  \\ R^+ & T^+ \end{bmatrix}\,
 \begin{bmatrix} F^- \\ F^+ \end{bmatrix} ,
 \end{equation}
where the transmission operators are
 \begin{subequations}
 \begin{eqnarray}
 \labe{Tminus} T^- &=& i\, t\, t_s \left(C^- C^+\right)^{1/2} S\, A, \nd \\
 \labe{Tplus} T^+ &=& i\, t\, t_s\, S \left(C^+ C^-\right)^{1/2} A
 ,
 \end{eqnarray}
 \end{subequations}
and the reflection operators are
 \begin{subequations}
 \begin{eqnarray}
 \labe{Rminus} R^- &=& -r\, e^{+i\,2 k\, \Delta z} \, C^-\, A , \nd \\
 \labe{Rplus} R^+ &=& -r\, t_s^2\, e^{-i\,2 k\, \Delta z} \, S\, C^+\, S\, A
 .
 \end{eqnarray}
 \end{subequations}
In the lossless case where $r^2 + t^2 = t^2_s = 1$ and $A$ is the
identity matrix, the explicit unitarity of the $C^+$, $C^-$, and
$S$ matrix operators guarantees that $|E^+|^2 + |E^-|^2 = |F^+|^2
+ |F^-|^2$.

 \epsx{2.75in}{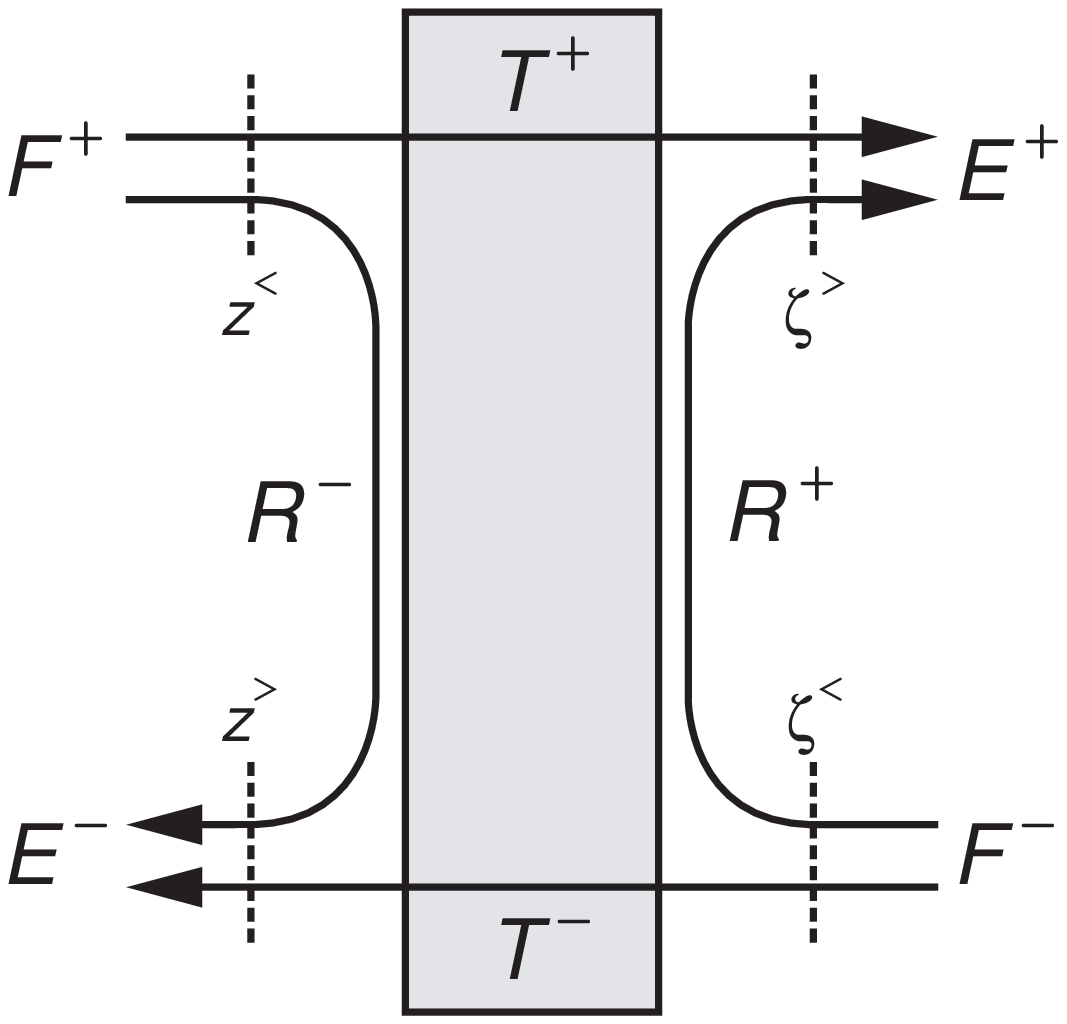}{Schematic diagram of the generalized
mirror transfer matrix given by \eqr{mxfer}. The HR-coated surface
of the mirror is on the left. The labels $z^<$ and $z^>$ denote
the input and output reference planes, respectively, near the HR
surface, while $\zeta^<$ and $\zeta^>$ are the input and output
reference planes near the AR surface.}

 \subsection{Beamsplitter\labs{beamsplitter}}

 \subsubsection{Aperture Diffraction and Thermal Focusing Operators}
The analysis of the aperture diffraction and thermal focusing
operators for the case of a 45$^\circ$ beamsplitter is similar to
that of a mirror. However, even though the beamsplitter has the
same cylindrically symmetric substrate as the mirror shown in
\fig{mirror_hv}, the non-normal angle of incidence will require
the elements of the perturbative matrix operators to be computed
numerically. We will assume here that the HR-coated surface of the
beamsplitter is flat, allowing us to ignore both curvature
mismatch and astigmatism when we build the transfer matrix in
\sct{bsxfer}.

The aperture diffraction operator for \emph{reflection} from the
HR-coated surface of the beamsplitter is given by
\begin{equation} \labe{ABSdef}
A_{m n, m^\prime n^\prime} = \iint_{\mathcal{E}} dx\,dy\;
u^\dagger_{m n}(x, y)\, u_{m^\prime n^\prime}(x, y) ,
\end{equation}
where $\mathcal{E}$ represents an elliptical aperture with
semimajor ($y$) axis $a$ and semiminor ($x$) axis $a/\sqrt{2}$.
Strictly speaking, the diffraction operator for
\emph{transmission} through the substrate is more complicated than
\eqr{ABSdef} because of refraction at each vacuum-dielectric
interface, particularly since the effective clear aperture (in the
horizontal plane) after propagation through the substrate has been
reduced by the distance $h/2 \eta$, or about 1.4~cm for LIGO. In
principle, we can construct separate operators for reflective and
transmissive aperture diffraction, but in practice we have found
that the resulting effect on the recycled power enhancement of the
standard PRFPMI configuration shown in \fig{caltech_prifo} was
negligible. Hence, in our simulations, we have made the
simplifying approximation that \eqr{ABSdef} can be used to
represent aperture diffraction effects for both reflection and
transmission.

The distribution of absorbed power within the beam\-split\-ter
volume is illustrated in \fig{bs_abs}, where we have introduced
the primary propagation paths labeled parallel ($\parallel$) and
perpendicular ($\perp$) relative to the propagation of the input
laser field in \fig{caltech_prifo}. Following the conventions of
\sct{mir_thermal_lens} as closely as possible, we then label the
input fields incident on the front (HR-coated) side of the mirror
with $\operatorname{sign}(\mathbf{\hat{k} \cdot \hat{z}}) > 0$ as
$F^{+\parallel}$ and $F^{+\perp}$, and those incident on the back
(AR-coated) side of the mirror, with
$\operatorname{sign}(\mathbf{\hat{k} \cdot \hat{z}}) < 0$ as
$F^{-\parallel}$ and $F^{-\perp}$. Together with the four
corresponding output fields shown in \fig{bs_abs}, these fields
generate five distinct optical power absorption regions. The total
TEM$_{00}$ optical power absorbed in each of these regions is
listed in \tbl{bsabs}, where $h^\prime \equiv h/\sqrt{1 - 1/2
\eta^2 }$ is the distance traveled through the substrate by the
refracted beams.

\epsx{2.75in}{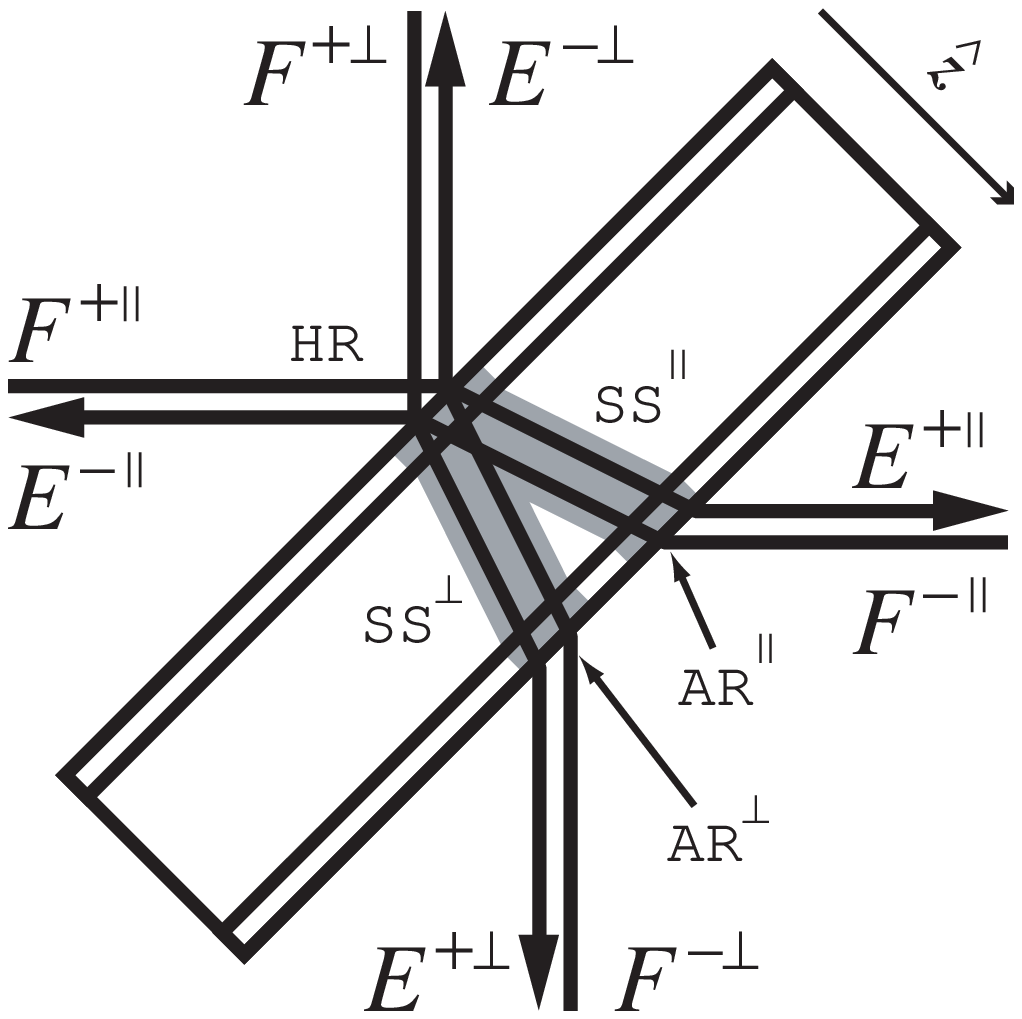}{Interaction of the two primary propagation
paths
--- parallel ($\parallel$) and perpendicular ($\perp$) ---
with the five optical absorption regions that cause thermal
focusing in the beamsplitter substrate
--- SS$^\parallel$, SS$^\perp$, HR, AR$^\parallel$, and AR$^\perp$.}

\begin{table}
\caption{Total absorbed TEM$_{00}$ powers for each of the five
absorption regions shown in \fig{bs_abs}, summed over the
lowest-order transverse modes of all active sidebands. The factors
of $\half$ arise from our normalization convention for the laser
electric field.\labt{bsabs}}
\begin{ruledtabular}
\begin{tabular}{cl}
Region & \multicolumn{1}{c}{Absorbed Power} \\ \hline
 SS$^\parallel$   &   $P^\parallel_s = \alpha_s h^\prime \, \half \sum_q \left(\left|E^{+\parallel}_{00q}\right|^2 + \left|F^{-\parallel}_{00q}\right|^2\right)$
\\
 SS$^\perp$      & $P^\perp_s = \alpha_s h^\prime \, \half \sum_q \left(\left|E^{+\perp}_{00q}\right|^2 + \left|F^{-\perp}_{00q}\right|^2\right)$
\\
 HR      & $P_h = \ensuremath{a_\mathit{hr}}\, \half \sum_q \biggl(\left|F^{+\parallel}_{00q}\right|^2 +
\left|F^{-\parallel}_{00q}\right|^2 $ \\ &\qquad\qquad\qquad $+
\left|F^{+\perp}_{00q}\right|^2 +
\left|F^{-\perp}_{00q}\right|^2\biggr)$ \\
 AR$^\parallel$    & $P^\parallel_a = \ensuremath{a_\mathit{ar}}\, \half \sum_q \left(\left|E^{+\parallel}_{00q}\right|^2 +
\left|F^{-\parallel}_{00q}\right|^2\right)$ \\
 AR$^\perp$     & $P^\perp_a = \ensuremath{a_\mathit{ar}}\, \half \sum_q \left(\left|E^{+\perp}_{00q}\right|^2 +
\left|F^{-\perp}_{00q}\right|^2\right)$ \\
\end{tabular}
\end{ruledtabular}
\end{table}

Diffusion of heat from each of the optical absorption regions
contributes to the total OPD distortion for both the parallel and
perpendicular propagation paths. Following the conventions we
established in \sct{mir_thermal_lens} and \eqs{Phidef}, for the
\emph{parallel} propagation path shown in \fig{bs_abs}, we denote
the substrate OPD phase operators corresponding to the five
absorption regions SS$^\parallel$, SS$^\perp$, HR, AR$^\parallel$,
and AR$^\perp$ as $\Phi^\parallel_s$, $\Phi^\perp_s$,
$\Phi^\prime_c$, $\Phi^\parallel_c$, and $\Phi^\perp_c$,
respectively. Since the perpendicular and parallel propagation
paths are symmetric under reflection in the $x$-$z$ plane about
the $z$ axis, the same set of operators can be used to represent
the substrate OPD distortion for perpendicular propagation, but in
the order $\Phi^\perp_s$, $\Phi^\parallel_s$, $\Phi^\prime_c$,
$\Phi^\perp_c$, and $\Phi^\parallel_c$. Hence, in a particular
basis, the net substrate thermal OPD matrix operators for the two
propagation paths can be written using our unitary approximation
as
 \begin{subequations}
 \labe{Sbs}
 \begin{eqnarray}
 S^\parallel &=& \exp \biggl\{i\, \biggl[P^\parallel_s\,
\Phi^\parallel_s + P^\perp_s\, \Phi^\perp_s \nonumber \\ &\qquad&
+ P_h\, \Phi^\prime_c + P^\parallel_a\, \Phi^\parallel_c +
P^\perp_a\, \Phi^\perp_c \biggr]\biggr\}, \nd \\
 S^\perp &=& \exp \biggl\{i\,
\biggl[P^\parallel_s\, \Phi^\perp_s+ P^\perp_s\, \Phi^\parallel_s
\nonumber \\ &\qquad& + P_h\, \Phi^\prime_c + P^\parallel_a\,
\Phi^\perp_c + P^\perp_a\, \Phi^\parallel_c \biggr]\biggr\}.
\end{eqnarray}
\end{subequations}

For the purposes of the initial LIGO simulations discussed in
\sct{ifo_sim}, we first used finite-element software to compute
the temperature distribution everywhere in the beamsplitter
substrate due to absorption of 1 Watt of TEM$_{0 0}$ optical power
(choosing the initial LIGO spot size at the beamsplitter) in each
of the five regions shown in \fig{bs_abs}. Following the same
basic approach as that of \sct{mir_thermal_lens}, we then computed
the matrix elements of the five parallel OPD operators in
\eqr{Sbs} using a Hermite-Gauss basis having the same spot size.
In our simulation runs, we simply scaled each matrix by the
appropriate absorbed power, and then computed $S^\parallel$ and
$S^\perp$ by matrix exponentiation.

 \subsubsection{Transfer Matrix\labs{bsxfer}}
As in the case of the mirror described in \sct{mxfer}, we can
construct a generalized transfer matrix for the beamsplitter that
includes the effects of aperturing, thermal focusing, and
longitudinal microdisplacement. Once again, we adopt the
industry-standard convention for the phases of the quarter-wave
amplitude reflection and transmission coefficients of the HR
coating, and we apply the substrate/AR amplitude transmission
coefficient given by \eqr{mirsubtrans} with $h \rightarrow
h^\prime$. However, as shown in \sct{MIpseudolocker}, the
definition of the beamsplitter position is more subtle than that
of the mirror: it consists of a static contribution $z$ that
represents the location chosen for the beamsplitter when used in
cold, unperturbed systems, and a dynamic contribution $\Delta z$
that is used in heated interferometers to maintain a given
intracavity phase condition. Hence, if the beamsplitter in
\fig{bs_abs} is displaced in the positive $\mathbf{\hat{z}}$
direction (i.e., away from the HR coating along the beamsplitter
axis of cylindrical symmetry) by the total displacement $z +
\Delta z$, then reflection from the ``front'' of the beamsplitter
introduces an additional relative phase $\exp[+i 2 \sqrt{2} k (z +
\Delta z)]$, while reflection from the ``back'' of the
beamsplitter requires a phase adjustment $\exp[-i 2 \sqrt{2} k (z
+ \Delta z)]$.

Collecting these results, the final beamsplitter transfer matrix
depicted in \fig{bs_xfer} is
 \begin{equation} \labe{bxfer}
 \begin{bmatrix} E^{-\parallel} \\ E^{+\perp} \\ E^{+\parallel} \\ E^{-\perp} \end{bmatrix} =
 \begin{bmatrix} T^{-\parallel} & R^{-\parallel} & 0 & 0 \\ R^{+\perp} & T^{+\perp} & 0 & 0 \\
                 0 & 0 & T^{+\parallel} & R^{+\parallel} \\ 0 & 0 & R^{-\perp} & T^{-\perp} \\\end{bmatrix}\,
 \begin{bmatrix} F^{-\parallel} \\ F^{+\perp} \\ F^{+\parallel} \\ F^{-\perp} \end{bmatrix} ,
 \end{equation}
where, in the absence of curvature mismatch, the transmission
operators are
 \begin{subequations}
 \begin{eqnarray}
 \labe{Tpar} T^{-\parallel} &=& T^{+\parallel} = i\, t\, t^\prime_s\, S^\parallel A, \nd \\
 \labe{Tperp} T^{-\perp} &=& T^{+\perp} = i\, t\, t^\prime_s\, S^\perp A
 ,
 \end{eqnarray}
 \end{subequations}
 and the reflection operators are
 \begin{subequations} \labe{bref}
 \begin{eqnarray}
 \labe{Rbplus} R^{-\parallel} &=& R^{-\perp} = -r\, e^{+i\,\sqrt{2} k\, (z + \Delta z)} \, A , \\
 \labe{Rbminuspar} R^{+\parallel} &=& -r\, t_s^{\prime\, 2}\, e^{-i\,\sqrt{2} k\, (z + \Delta z)}
  \, S^\parallel\, S^\perp\, A , \nd \\
 \labe{Rbminusperp} R^{+\perp} &=& -r\, t_s^{\prime\, 2}\, e^{-i\,\sqrt{2} k\, (z + \Delta z)}
  \, S^\perp\, S^\parallel\, A .
 \end{eqnarray}
 \end{subequations}
In the lossless case where $r^2 + t^2 = t^{\prime\, 2}_s = 1$ and
$A = 1$, the explicit unitarity of the $S^\parallel$ and $S^\perp$
matrix operators guarantees that $|E^{+\perp}|^2 +
|E^{-\parallel}|^2 = |F^{+\perp}|^2 + |F^{-\parallel}|^2$ and
$|E^{+\parallel}|^2 + |E^{-\perp}|^2 = |F^{+\parallel}|^2 +
|F^{-\perp}|^2$.

\begin{figure}
   \centering
   \subfigure[Cross-coupling for $+\perp$ and $-\parallel$]{\labf{bs_xfer_a}
 \includegraphics[width=3.25in]{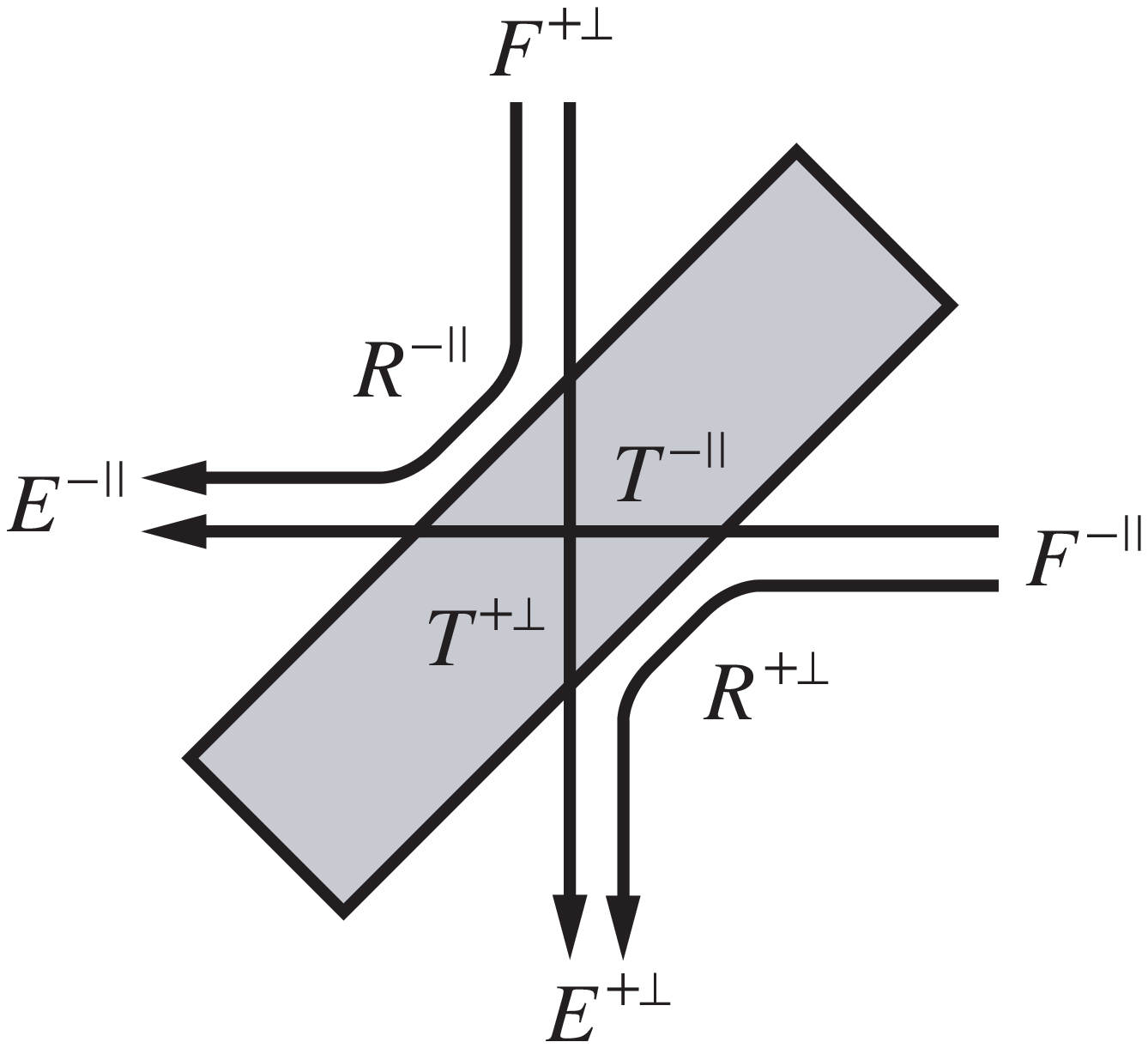}}\quad
   \subfigure[Cross-coupling for $-\perp$ and $+\parallel$]{\labf{bs_xfer_b}
 \includegraphics[width=3.25in]{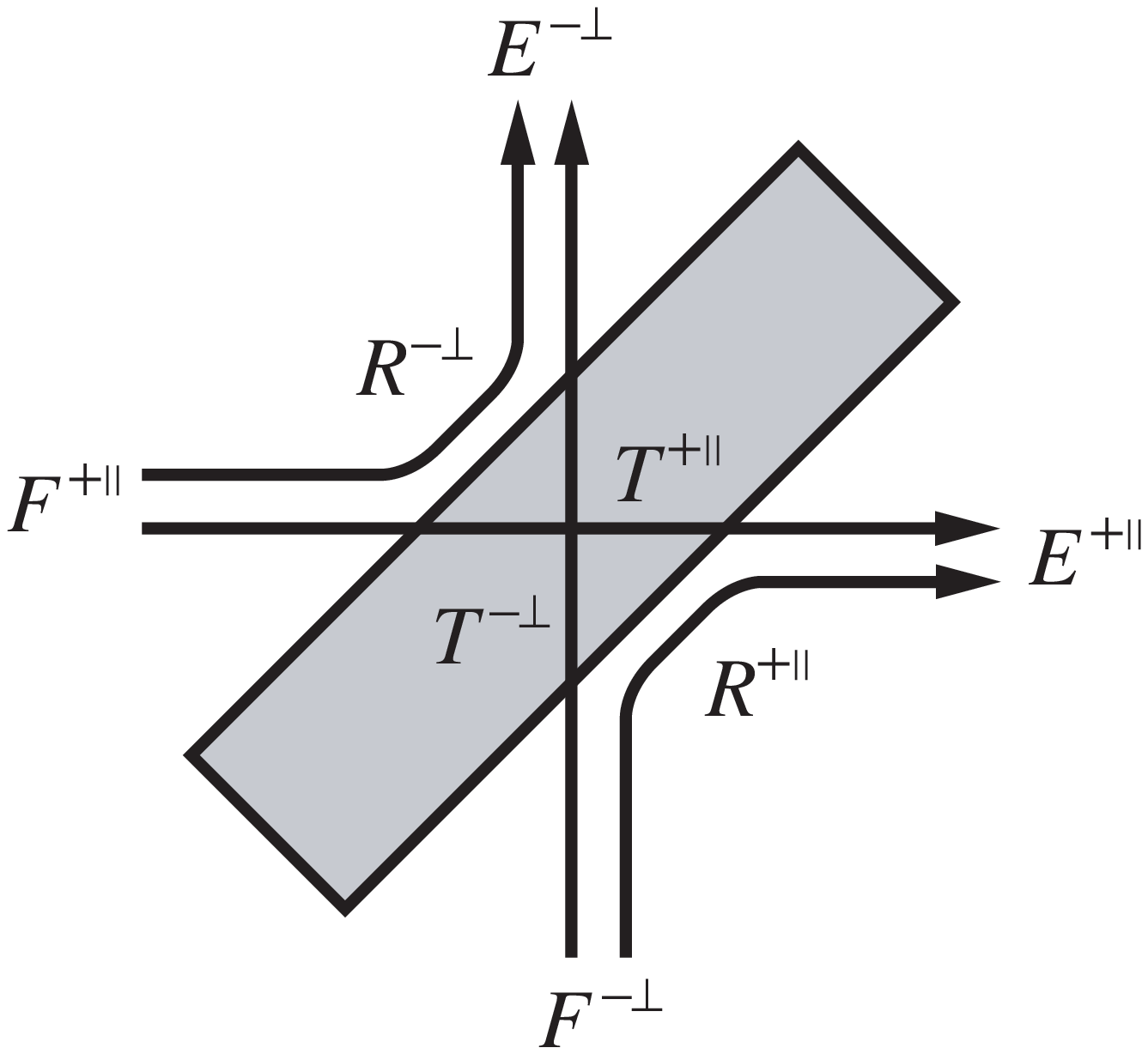}}
   \caption{\labf{bs_xfer} Schematic diagram of the generalized
beamsplitter transfer matrix given by \eqr{bxfer}. The HR surface
of the mirror is on the upper left.}
 \end{figure}

\subsection{Fabry-Perot Interferometers\labs{FPI}}
The propagation schematic diagram of a Fabry-Perot interferometer
(FPI) is shown in \fig{fpi_xfer}. The front (HR) input and output
ports of mirror $\mathcal{M}_1$ and $\mathcal{M}_3$ are coupled
through a propagation distance $L_{13}$. We can simplify later
calculations of the optical performance of a power-recycled
gravitational-wave interferometer if we first determine a transfer
matrix for the FPI, and then establish a procedure maintaining an
optimum level of intracavity field amplitude enhancement in the
presence of systematic perturbations. Although we develop the FPI
transfer matrix in a basis-independent fashion, and postpone the
discussion of the choice of an unperturbed basis set until we
calculate steady-state fields in \sct{IFOSS}, we note that we will
use the unperturbed Hermite-Gauss eigenmodes of one of the
Fabry-Perot interferometers in \fig{caltech_prifo} to develop a
basis set for the entire gravitational-wave detector.

\subsubsection{Transfer Matrix}
\epsw{5.0in}{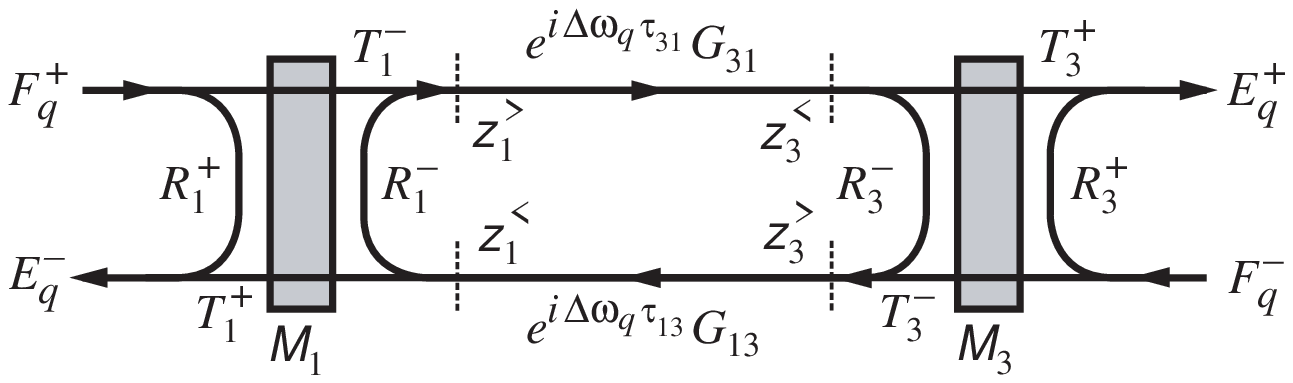}{Schematic diagram of the enhancement,
reflection, and transmission operators of a Fabry-Perot
interferometer, defined by \eqr{HFPI}, \eqr{RFPI}, and \eqr{TFPI},
respectively.}

Consider the schematic representation of the enhancement,
reflection, and transmission operators of the Fabry-Perot
interferometer shown in \fig{fpi_xfer}. If we express these
operators using the transfer matrices of mirrors $\mathcal{M}_1$
and $\mathcal{M}_3$, described in \sct{mxfer}, then the
self-consistent forward propagation steps between the plane of
incidence of $\mathcal{M}_1$ at $z^<_1$ and the plane of incidence
of $\mathcal{M}_3$ at $z^>_3$ can be represented as
 \begin{subequations} \labe{fpienh}
 \begin{eqnarray}
 F^-_{1 q} &=& e^{i\, \domq\, \tau_{1 3}} G_{1 3} \left( R^-_3\, F^+_{3 q}
 + T^-_3\, F^-_q \right), \nd
 \\
 F^+_{3 q} &=& e^{i\, \domq\, \tau_{3 1}} G_{3 1} \left( R^-_1\, F^-_{1 q}
 + T^-_1\, F^+_q \right).
 \end{eqnarray}
 \end{subequations}
where $F^+_q$ is the amplitude of the external sideband field $q$
incident on $\mathcal{M}_1$, $F^-_q$ is the amplitude of the
external sideband field $q$ incident on $\mathcal{M}_3$, $F^-_{1
q}$ is the amplitude of frequency component $q$ incident on
reference plane $z^<_1$, and $F^+_{3 q}$ is the corresponding
field incident on reference plane $z^>_3$. Here $G_{31}$ is the
Gouy operator describing the forward propagation step from $z^>_1$
to $z^<_3$, while $G_{13}$ is the corresponding operator
describing propagation from $z^>_3$ to $z^<_1$. (In a
power-orthonormal basis, the matrices representing these operators
are identical.) If the length of the vacuum separating
$\mathcal{M}_1$ from $\mathcal{M}_3$ is $L_{31}$, then the
single-pass propagation time is $\tau_{31} \equiv L_{31}/c =
\tau_{13}$.

Using the concatenated column vectors $[E^-_{1 q}\quad E^-_{3
q}]^T$ and $[F^-_q\quad F^+_q]^T$, we can construct a system of
sparse matrix equations which can be efficiently solved
numerically using Gaussian elimination:
 \begin{equation} \labe{fpigeq}
 \left[1 - \hat{G}(\domq)\, \hat{R}^-\right] \begin{bmatrix} F^-_{1 q} \\ F^+_{3 q} \end{bmatrix} =
 \hat{G}(\domq)\, \hat{T}^-\,
 \begin{bmatrix} F^-_q \\ F^+_q \end{bmatrix} ,
 \end{equation}
 where
 \begin{equation} \labe{fpirthat}
 \hat{T}^- = \begin{bmatrix} T^-_3 & 0  \\ 0 & T^-_1
\end{bmatrix}
\quad\nd \quad \hat{R}^- = \begin{bmatrix} 0 & R^-_3 \\ R^-_1 & 0
\end{bmatrix}
\end{equation}
and
 \begin{equation} \labe{BFPI}
\hat{G}(\domg) = \begin{bmatrix} e^{i\, \domg\, \tau_{1 3}} G_{1
3} & 0
\\ 0 & e^{i\, \domg\, \tau_{3 1}} G_{3 1} \end{bmatrix}
  .
 \end{equation}
Alternatively, for a purely analytic calculation, we can define
the global enhancement operator
 \begin{equation} \labe{fpigloben}
 \hat{H}(\domg) = \left[1 - \hat{G}(\domg)\, \hat{R}^-\right]^{-1} \hat{G}(\domg)\,
 \hat{T}^-,
 \end{equation}
 and then solve \eqr{fpigeq} analytically to obtain
 \begin{equation} \labe{fpialen}
 \begin{bmatrix} F^-_{1 q} \\ F^+_{3 q} \end{bmatrix} =
 \begin{bmatrix} H^-_1(\domq) & H^+_1(\domq)  \\ H^-_3(\domq) & H^+_3(\domq) \end{bmatrix}\,
 \begin{bmatrix} F^-_q \\ F^+_q \end{bmatrix} ,
 \end{equation}
where the operators representing the enhancements at each of the
reference planes in \fig{fpi_xfer} are given by

 \begin{widetext}
 \begin{subequations} \labe{HFPI}
 \begin{eqnarray}
 H^-_1(\domg) &=& \left(1 - e^{i\, 2\, \domg\, \tau_{13}}
 G_{13}\, R^-_3\, G_{31}\,
R^-_1\right)^{-1} e^{i\, \domg\, \tau_{13}} G_{13}\, T^-_3 ,
\\
 \labe{Hp1} H^+_1(\domg) &=& \left(1 - e^{i\, 2\, \domg\, \tau_{13}}
 G_{13}\, R^-_3\, G_{31}\,
R^-_1\right)^{-1} e^{i\, 2\, \domg\, \tau_{13}} G_{13}\, R^-_3\,
G_{31}\, T^-_1 , \\
 H^-_3(\domg) &=& \left(1 - e^{i\, 2\, \domg\, \tau_{31}}
 G_{31}\, R^-_1\, G_{13}\,
R^-_3\right)^{-1} e^{i\, 2\, \domg\, \tau_{31}} G_{31}\, R^-_1\,
G_{13}\, T^-_3 , \nd \\
 H^+_3(\domg) &=& \left(1 - e^{i\, 2\, \domg\, \tau_{31}}
 G_{31}\, R^-_1\, G_{13}\,
R^-_3\right)^{-1} e^{i\, \domg\, \tau_{31}} G_{31}\, T^-_1 .
 \end{eqnarray}
 \end{subequations}
 \end{widetext}

From \fig{fpi_xfer}, we see that each of the output fields $E^-_q$
and $E^+_q$ can be expressed as the sum of a prompt reflection of
the corresponding input field and an enhanced intracavity field
transmitted through the adjacent mirror, giving
 \begin{subequations} \labe{fpiout}
 \begin{eqnarray}
 E^-_q &=&  T^+_1\, F^-_{1 q} + R^+_1\, F^+_q , \nd \\
 E^+_q &=& T^+_3\, F^+_{3 q} + R^+_3\, F^-_q ,
 \end{eqnarray}
 \end{subequations}
 or, from \eqr{fpigloben} and \eqr{fpialen},
 \begin{equation} \labe{fpixfr}
 \begin{bmatrix} E^-_q \\ E^+_q \end{bmatrix} =
 \left[ \hat{R}^+ + \hat{T}^+\, \hat{H}(\domq)\right]\,
 \begin{bmatrix} F^-_q \\ F^+_q \end{bmatrix} ,
 \end{equation}
 where
 \begin{equation} \labe{fpirthatp}
 \hat{T}^+ = \begin{bmatrix} T^+_1 & 0  \\ 0 & T^+_3
\end{bmatrix}
\quad\nd \quad \hat{R}^+ = \begin{bmatrix} 0 & R^+_1 \\ R^+_3 & 0
\end{bmatrix} .
\end{equation}
We can now construct an explicit transfer matrix similar to
\eqr{mxfer} for the FPI as a monolithic optical element. Using
\fig{mirror_xfer} as a guide, \eqr{fpixfr} gives:
 \begin{equation} \labe{fpixfer}
 \begin{bmatrix} E^-_q \\ E^+_q \end{bmatrix} =
 \begin{bmatrix} T^-_{\text{FPI}}(\domq) & R^-_{\text{FPI}}(\domq)  \\ R^+_{\text{FPI}}(\domq) & T^+_{\text{FPI}}(\domq) \end{bmatrix}\,
 \begin{bmatrix} F^-_q \\ F^+_q \end{bmatrix} ,
 \end{equation}
where the transmission operators are
 \begin{subequations} \labe{TFPI}
 \begin{eqnarray}
 \labe{Tfpiminus} T^-_{\text{FPI}}(\domg) &=& T^+_1\, H^-_1(\domg), \nd \\
 \labe{Tfpiplus} T^+_{\text{FPI}}(\domg) &=& T^+_3\, H^+_3(\domg) ,
 \end{eqnarray}
 \end{subequations}
and the reflection operators are
 \begin{subequations} \labe{RFPI}
 \begin{eqnarray}
 \labe{Rfpiminus} R^-_{\text{FPI}}(\domg) &=& R^+_1 + T^+_1\, H^+_1(\domg), \nd \\
 \labe{Rfpiplus} R^+_{\text{FPI}}(\domg) &=& R^+_3 + T^+_3\, H^-_3(\domg)
 .
 \end{eqnarray}
 \end{subequations}

\subsubsection{Idealized Simulation of
Servo-Controlled Resonator Length Locking\labs{FPIpseudolocker}}
Suppose that the FPI has been initialized using a particular set
of configuration parameters for mirrors $\mathcal{M}_1$ and
$\mathcal{M}_3$, the initial thermal loads are negligible, and the
initial microdisplacements of both mirrors are set to zero. At
some later time, a perturbation is introduced, such as a different
curvature for $\mathcal{M}_1$, that alters the resonance condition
of the cavity. We assume that a biorthonormal set of unperturbed
basis functions $u_{m n}({\bf r})$ has been chosen to represent
both the intracavity and extracavity transverse laser fields prior
to any perturbations, and we wish to compute the corresponding
perturbed transfer mirror matrix given by \eqr{mxfer} in that
unperturbed basis.

In any experiment using an FPI, the fields emerging from the
resonator will be monitored and the cavity length will be adjusted
to maintain optimum enhancement. However, rather than simulate a
servo-mechanical control loop, we choose to compute a
microdisplacement $\Delta z_1$ for $\mathcal{M}_1$ directly from a
numerical determination of the net round-trip phase accumulated by
the lowest-loss carrier eigenmode of the resonator. Beginning and
ending at the reference plane $z^<_1$, the perturbed round-trip
propagator at the carrier frequency for the FPI shown in
\fig{fpi_xfer} is
\begin{equation}
K_{\text{FPI}}(\Delta z_1) = e^{-i\, 2\, k\, \Delta z_1}
 G_{13}\, R^-_3\, G_{31}\, R^-_1
\end{equation}
where we have chosen to extract from \eqr{Rminus} the explicit
phase offset arising from a nonzero value of $\Delta z_1$.

We can then compute the spectrum of eigenvalues of
$K_{\text{FPI}}(0)$, and select the eigenvalue $\lambda_{0} \equiv
|\lambda_{0}| e^{i\, \phi_{0}}$ with the largest magnitude (i.e.,
the smallest round-trip loss). Clearly, the corresponding
eigenvector $E_{0}$ is also an eigenvector of the matrix operator
$K_{\text{FPI}}(\Delta z_1)$, with the eigenvalue $\lambda_1 =
\lambda_{0} e^{-i\, 2\, k\, \Delta z_1}$. Hence, if we choose
\begin{equation}
\Delta z_1 = \frac{\phi_{0}}{2\,k} ,
\end{equation}
then the net round-trip phase accumulated by $E_{0}$ will be zero,
and that mode will satisfy the resonance condition. In our
simulations, we maintain a record of the most recently chosen
lowest-loss eigenvector, and in the case of degeneracy we choose
the eigenmode having the largest value of the inner product with
the previous eigenvector.

Suppose that we have followed this procedure and computed the
eigenvector $E_0$ of the matrix operator
$K_{\text{FPI}}(\phi_0/2\,k)$, representing a maximally resonant
field amplitude at reference plane $z^<_1$ in \fig{fpi_xfer}. We
can mode-match this field by choosing input fields that optimally
couple to $E_0$ through the global enhancement operator
$\hat{H}(0)$. For example, if we solve the matrix equation $E_0 =
H^+_1(0)\, F^+_0$ for $F^+_0$, where $H^+_1(0)$ is given by
\eqr{Hp1}, we obtain the intuitively self-consistent result
 \begin{equation} \labe{fpimm}
 F^+_0 = \left(\frac{1 - \left|\lambda_0\right|}{\lambda^*_0}\right)\,
 \left(T^-_1\right)^{-1} R^-_1\, E_0 .
 \end{equation}
A similar mode-matching condition for the back reference plane and
incident field can be found using the same procedure.

\subsection{Michelson Interferometer}
The propagation schematic of a generalized Michelson
interferometer (MI) is shown in \fig{mi_xfer}. The back (AR)
parallel input and output ports are coupled to either a mirror or
FPI $\mathcal{M}_\text{I}$ through a propagation distance
$L_{16}$, and the front (HR) perpendicular ports are coupled to
either a mirror or FPI $\mathcal{M}_\text{II}$ through a
propagation distance $L_{26}$. We can simplify later calculations
of the optical performance of a power-recycled gravitational wave
interferometer if we first determine a transfer matrix for the MI,
and then establish a procedure for satisfying the dark port
condition in the presence of perturbations. As in \sct{FPI}, we
develop the MI transfer matrix in a basis-independent fashion, and
postpone the discussion of the choice of an unperturbed basis set
until we calculate steady-state fields in \sct{IFOSS}.

\epsf{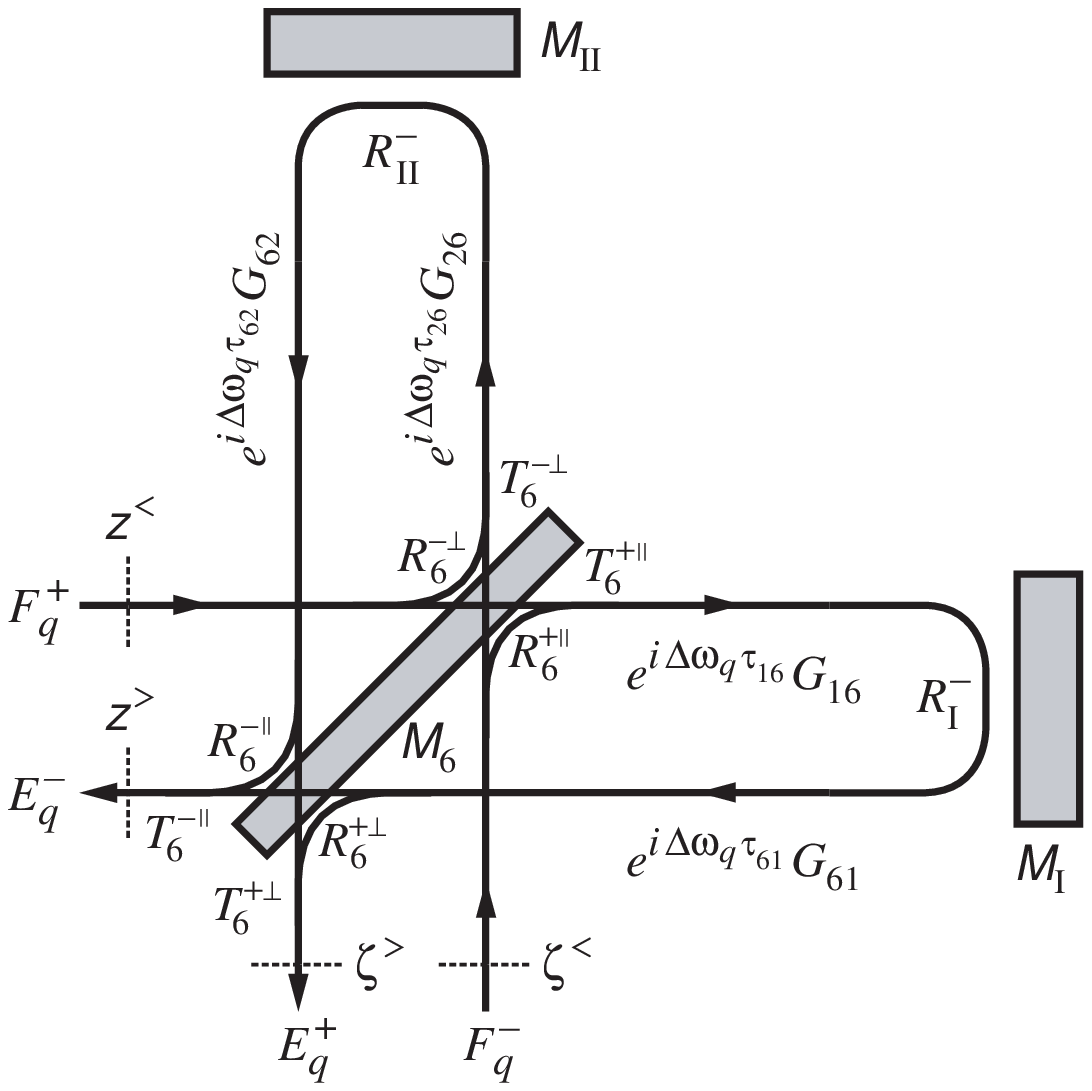}{Schematic diagram of the reflection and
transmission operators of a Michelson interferometer, defined by
\eqr{MIxfer}. Note that the internal reflection operators
$R^-_\text{I}(\domg_q)$ and $R^-_\text{II}(\domg_q)$ may each
represent either the reflection operator of a mirror, given by
\eqr{Rminus}, or that of a Fabry-Perot interferometer, given by
\eqr{Rfpiminus}. The labels $z^<$ and $z^>$ denote the input and
output reference planes, respectively, near the HR surface of the
beamsplitter , while $\zeta^<$ and $\zeta^>$ are the input and
output reference planes near the BS AR surface.}

 \subsubsection{Transfer Matrix\labs{mixfer}}
By comparing \fig{bs_xfer} and \fig{mi_xfer}, we can connect the
MI input and output fields of frequency component $q$ with the
front parallel and back perpendicular fields of the beamsplitter
using the associations
 \begin{alignat*}{2}
 E^{-\parallel} &\longrightarrow E^-_q , & \qquad F^{-\perp} &\longrightarrow
 F^+_q\\
 E^{+\perp} &\longrightarrow E^+_q & \qquad F^{+\parallel} &\longrightarrow
 F^-_q
 \end{alignat*}
As shown in \fig{mi_xfer}, the back parallel fields are coupled
through reflection from optical component $\mathcal{M}_\text{I}$,
and the front perpendicular fields are coupled through reflection
from $\mathcal{M}_\text{II}$. Therefore, for frequency component
$q$, we have
 \begin{subequations} \labe{mi_coupling}
 \begin{eqnarray}
 F^{-\parallel}_q &=& K_{616}(\domg_q)\, E^{+\parallel}_q , \nd \\
 F^{+\perp}_q &=& K_{626}(\domg_q)\, E^{-\perp}_q .
 \end{eqnarray}
\end{subequations}
where
\begin{subequations}
\begin{eqnarray}
\labe{K616} K_{616}(\domg) &\equiv& e^{i\, 2\, \domg\,
\tau_{61}}\, G_{61}\, R^-_\text{I}(\domg)\, G_{16} , \nd \\
 \labe{K626} K_{626}(\domg) &\equiv& e^{i\, 2\, \domg\,
\tau_{62}}\, G_{62}\, R^-_\text{II}(\domg)\, G_{26}.
\end{eqnarray}
\end{subequations}
In \eqr{K616}, $G_{16}$ is the Gouy operator describing the
forward propagation step from the back parallel output port of the
beamsplitter $\mathcal{M}_6$ to the input port of
$\mathcal{M}_\text{I}$, while $G_{61}$ is the corresponding
operator describing propagation from the output port of
$\mathcal{M}_\text{I}$ to the back parallel input port of
$\mathcal{M}_6$. If the length of the vacuum separating
$\mathcal{M}_6$ from $\mathcal{M}_\text{I}$ is $L_{16}$, then the
single-pass propagation time is $\tau_{16} \equiv L_{16}/c =
\tau_{61}$. In \eqr{K626}, there is an equivalent set of
propagation operators and times, with $1 \rightarrow 2$ and
$\text{I} \rightarrow \text{II}$. The internal reflection
operators $R^-_\text{I}(\domg)$ and $R^-_\text{II}(\domg)$ may
each represent either the reflection operator of a mirror, given
by \eqr{Rminus}, or that of a Fabry-Perot interferometer, given by
\eqr{Rfpiminus}.

By substituting \eqr{mi_coupling} into \eqr{bxfer}, we can
immediately construct a transfer matrix similar to \eqr{mxfer} for
the MI as a monolithic optical element. Using \fig{mirror_xfer} as
a guide, we find:
 \begin{equation} \labe{mixfer}
 \begin{bmatrix} E^-_q \\ E^+_q \end{bmatrix} =
 \begin{bmatrix} T^-_{\text{MI}}(\domg_q) & R^-_{\text{MI}}(\domg_q)  \\ R^+_{\text{MI}}(\domg_q) & T^+_{\text{MI}}(\domg_q) \end{bmatrix}\,
 \begin{bmatrix} F^-_q \\ F^+_q \end{bmatrix} ,
 \end{equation}
where the individual reflection and transmission operators are
defined by the matrix product
\begin{equation} \labe{MIxfer}
\begin{split}
 &\begin{bmatrix} T^-_{\text{MI}}(\domg) & R^-_{\text{MI}}(\domg)  \\ R^+_{\text{MI}}(\domg) & T^+_{\text{MI}}(\domg) \end{bmatrix} =
  \begin{bmatrix} T^{-\parallel}_6 & R^{-\parallel}_6  \\ R^{+\perp}_6 & T^{+\perp}_6
 \end{bmatrix} \\ &\times\, \begin{bmatrix} K_{616}(\domg)\negthickspace & \negthickspace 0  \\ 0\negthickspace & \negthickspace K_{626}(\domg) \end{bmatrix}
    \begin{bmatrix} T^{+\parallel}_6 & R^{+\parallel}_6  \\ R^{-\perp}_6 & T^{-\perp}_6 \end{bmatrix}
    \begin{bmatrix} 0 & 1 \\ 1 & 0 \end{bmatrix}.
 \end{split}
\end{equation}

 \subsubsection{Idealized Simulation of the Dark Port
 Condition\labs{MIpseudolocker}}

Interferometric gravitational-wave Michelson interferometers
typically operate with the beamsplitter positioned so that the
back perpendicular (antisymmetric) output port generates a dark
fringe. In the model MI developed in the previous section, we can
adjust the static position $z_6$ of the beamsplitter to ensure
that an unperturbed, mode-matched fundamental field (i.e., $\{m,
n, q\} = \{0, 0, 0\}$) incident on the beamsplitter's front
parallel input port is cancelled at the back perpendicular output
port. Given our conventions for the phases of the amplitude
reflection and transmission coefficients of a high-reflectance
coating, we choose $z_6 = -\pi/2 \sqrt{2} k$\cite{bea99} and
obtain from \eqr{bref}
 \begin{subequations} \labe{MIR}
 \begin{eqnarray}
 \labe{MIRbplus} R^{-\parallel}_6 &=& R^{-\perp}_6 = i\, r_6\, e^{+i\,\sqrt{2} k\, \Delta z_6} \, A_6 , \\
 \labe{MIRbminuspar} R^{+\parallel}_6 &=& -i\, r_6\, t_s^2\, e^{-i\,\sqrt{2} k\, \Delta z_6}
  \, S^\parallel_6\, S^\perp_6\, A_6 , \nd \\
 \labe{MIRbminusperp} R^{+\perp}_6 &=& -i\, r_6\, t_s^2\, e^{-i\,\sqrt{2} k\, \Delta z_6}
  \, S^\perp_6\, S^\parallel_6\, A_6 .
 \end{eqnarray}
 \end{subequations}
If a compensation plate is placed in the vacuum between the
beamsplitter and $\mathcal{M}_\text{II}$, then (depending upon its
orientation angle) it can be represented as either a mirror or a
beamsplitter substrate with two antireflecting coatings and no
curvature, and the corresponding operator can be inserted into the
perpendicular propagator $K_{626}(\domg)$ given by \eqr{K626}.

Suppose that the MI has been initialized using a particular set of
configuration parameters, and in the absence of a thermal lens in
any substrate the dynamic microdisplacements of the beamsplitter
and the mirrors are set to zero. If the static beamsplitter
position has been set to satisfy the dark port condition for $q =
0$, then an unperturbed TEM$_{00}$ field incident on the front
parallel port produces a vanishingly small output at the back
perpendicular port. Insofar as it is possible, we wish to develop
an algorithm that maintains this dark port condition when the
system has been perturbed.

Consider an arbitrary field $F^+_0$ incident on the front input
port of the MI. At the back output port, the transmitted field can
be found by simplifying \eqr{mixfer} for $q = 0$, giving
 \begin{equation}
 E^+_0 = \left[D_1(\Delta z_6) + D_2(\Delta z_6)\right]\,
 F^+_0 ,
 \end{equation}
 where we choose to extract from \eqr{MIR} the explicit phase
offset arising from a nonzero value of $\Delta z_6$, and write
 \begin{subequations}
 \begin{eqnarray}
 D_1(\Delta z_6) &=& e^{+i\, \sqrt{2}\, k\, \Delta z_6}\,
R^{+\perp}_6\, K_{616}(0)\, T^{+\parallel}_6 , \nd \\
 D_2(\Delta z_6) &=& e^{-i\, \sqrt{2}\, k\, \Delta z_6}\,
T^{+\perp}_6\, K_{626}(0)\, R^{-\perp}_6 .
 \end{eqnarray}
 \end{subequations}
The corresponding power measured by a square-law detector at the
dark port is therefore (within a constant)
\begin{equation} \labe{PMI}
\begin{split}
P^+_0 (\Delta z_6) &= \half\, E^{+\dag}_0\, E^+_0 \\
 &= \half\, F^{+\dag}_{0}\, D^\dag_1(\Delta\,
 z_6)\, D_2(\Delta z_6)\, F_{0} + \text{c.c.} \\
 &= \left|F^{+\dag}_{0}\, D^\dag_1(0)\, D_2(0)\, F_{0}\right|\,
  \\&\qquad \times\, \cos\left(2 \sqrt{2}\, k\, \Delta z_6 - \phi\right)
\end{split}
\end{equation}
where $\phi$ is the phase angle of $- F^{+\dag}_{0}\,
D^\dag_1(0)\, D_2(0)\, F^+_{0}$. In order to minimize $P^+_0
(\Delta z_6)$, we require that the condition $\partial\, P^+_0
(\Delta z_6)/\partial\, \Delta z_6 = 0$ be satisfied, or
\begin{equation} \labe{dz6}
\Delta z_6 = \frac{\phi}{2 \sqrt{2}\, k} .
\end{equation}
We will incorporate this procedure into a general algorithm for
simulating a servo-controlled resonator length-locking system for
an entire gravitational-wave interferometer in
\sct{IFOpseudolocker}.

\epsw{5.0in}{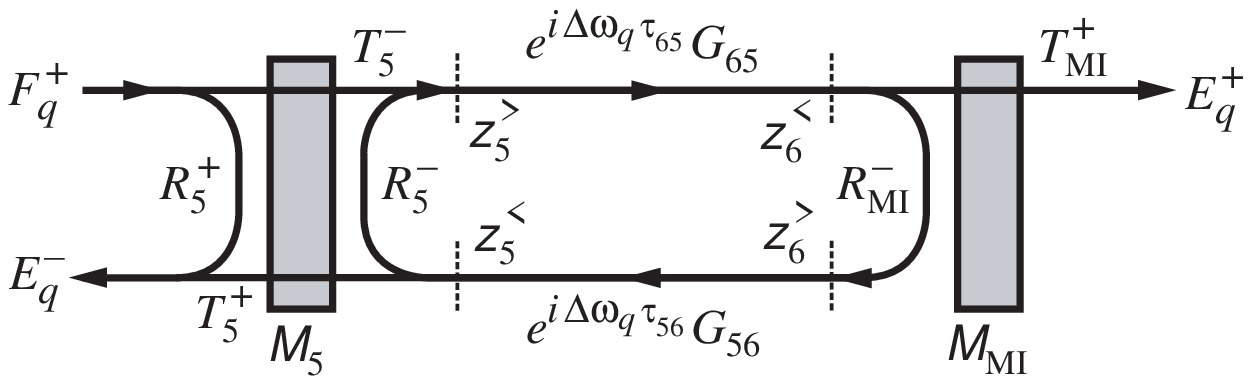}{Schematic diagram of the enhancement,
reflection, and transmission operators of a power-recycled
Fabry-Perot Michelson interferometer, defined by \eqr{HIFO} and
\eqr{TRIFO}, respectively.}

\section{Power-Recycled Fabry-Perot Michelson Interferometer\labs{prfpmi}}

In previous sections, we have constructed the transfer matrices of
the components needed to build a functionally complete description
of a power-recycled Fabry-Perot Michelson interferometer (PRFPMI).
The PRFPMI propagation schematic shown in \fig{prifo_xfer} is a
simplified form of the FPI schematic displayed in \fig{fpi_xfer},
and will allow us to determine the reduced transfer matrix with a
minimum of calculation. In this case, the mirror $\mathcal{M}_1$
is replaced by the power-recycling mirror $\mathcal{M}_5$, and the
mirror $\mathcal{M}_3$ is replaced by the Michelson interferometer
$\mathcal{M}_\text{MI}$. After determining the corresponding
enhancement, reflection, and transmission operators of the PRFPMI,
we develop a simulation of an idealized servo-controlled length
locking system that incorporates the dark port condition algorithm
described in \sct{MIpseudolocker}. We then demonstrate the
determination of a set of unperturbed basis functions that can be
used to describe the electromagnetic intracavity and extracavity
fields, and we develop a numerical algorithm for the computation
of the steady-state fields under the influence of both geometric
and thermal perturbations. After a detailed comparison of the
predictions of this model with those of a fast Fourier transform
code set in two important special cases, we evaluate the optical
performance of the baseline LIGO design.

 \subsection{Transfer Matrix\labs{ifo}}
Consider the schematic representation of the enhancement,
reflection, and transmission operators of the interferometer shown
in \fig{prifo_xfer}. If we express these operators using the
transfer matrices of $\mathcal{M}_1$ and $\mathcal{M}_\text{MI}$,
respectively described in \sct{mxfer} and \sct{mixfer}, then the
intracavity field enhancements can be represented as
 \begin{subequations} \labe{ifoenh}
 \begin{eqnarray}
 F^-_{5 q} &=& H^+_5(\domg_q) F^+_q , \nd \labe{E5qifo} \\
 F^+_{6 q} &=& H^+_6(\domg_q) F^+_q .
 \end{eqnarray}
 \end{subequations}
where $F^+_q$ is the amplitude of the external sideband field $q$
incident on $\mathcal{M}_5$, $F^-_{5 q}$ is the amplitude of
frequency component $q$ incident on reference plane $z^<_5$, and
$F^+_{6 q}$ is the corresponding field incident on reference plane
$z^<_6$. By comparison with \eqr{fpienh}, the enhancement
operators are
\begin{widetext}
\begin{subequations} \labe{HIFO}
 \begin{eqnarray}
 \labe{Hp5} H^+_5(\domg) &=& \left[1 - e^{i\, 2\, \domg\, \tau_{56}}
 G_{56}\, R^-_{\text{MI}}(\domg)\, G_{65}\,
R^-_5\right]^{-1} e^{i\, 2\, \domg\, \tau_{56}} G_{56}\,
R^-_{\text{MI}}(\domg)\, G_{65}\, T^-_5 , \\
 H^+_6(\domg) &=& \left[1 - e^{i\, 2\, \domg\, \tau_{65}}
 G_{65}\, R^-_5\, G_{56}\,
R^-_{\text{MI}}(\domg)\right]^{-1} e^{i\, \domg\, \tau_{65}}
G_{65}\, T^-_5 .
 \end{eqnarray}
 \end{subequations}
 \end{widetext}
where $G_{65}$ is the Gouy operator describing the forward
propagation step from $z^>_5$ to $z^<_6$, and $G_{56}$ is the
corresponding operator describing propagation from $z^>_6$ to
$z^<_5$. If the length of the vacuum separating $\mathcal{M}_5$
from $\mathcal{M}_\text{MI}$ is $L_{65}$, then the single-pass
propagation time is $\tau_{65} \equiv L_{65}/c = \tau_{56}$. The
front transmission and reflection operators of the power-recycling
mirror, $T^-_5$ and $R^-_5$, are given by \eqr{Tminus} and
\eqr{Rminus}, respectively. The front reflection operator of the
MI, $R^-_\text{MI}(\domg)$, is given by the appropriate element of
\eqr{MIxfer}.

Given the PRFPMI schematic shown in \fig{prifo_xfer} and the
enhancement operators $H^+_5(\domg)$ and $H^+_6(\domg)$, we obtain
the output fields $E^-_q$ and $E^+_q$ as
 \begin{subequations} \labe{prifoxfer}
 \begin{eqnarray}
 E^-_q &=& R^-_{\text{IFO}}(\domg_q) F^+_q , \nd \\
 E^+_q &=& T^+_{\text{IFO}}(\domg_q) F^+_q .
 \end{eqnarray}
 \end{subequations}
where the transmission and reflection operators are
 \begin{subequations} \labe{TRIFO}
 \begin{eqnarray}
 \labe{Tifoplus} T^+_{\text{IFO}}(\domg) &=& T^+_{\text{MI}}(\domg)\, H^+_6(\domg) , \nd \\
 \labe{Rifominus} R^-_{\text{IFO}}(\domg) &=& R^+_5 + T^+_5\, H^+_5(\domg)
 .
 \end{eqnarray}
 \end{subequations}

\subsection{Idealized Simulation of Servo-Controlled Resonator Length
Locking\labs{IFOpseudolocker}}

Although the transfer matrix schematics of the FPI and the PRFPMI
--- shown in \fig{fpi_xfer} and \fig{prifo_xfer}, respectively ---
are similar, the resonant structure of the MI and the dark port
condition significantly complicate simulations of an idealized
resonator length-locking system. Clearly, we need to adjust the
position of the power recycling mirror $\mathcal{M}_5$ to maintain
a ``round-trip'' resonance condition, but the round trip must
include both resonances in the Michelson arms and reflection from
a beamsplitter that has been infinitesimally displaced to force
compliance with the Michelson dark port condition. In fact, we
again choose to compute a microdisplacement $\Delta z_5$ for
$\mathcal{M}_5$ directly from a numerical determination of the net
round-trip phase accumulated by the lowest-loss carrier eigenmode
of the entire resonator. Beginning and ending at the reference
plane $z^<_5$, the perturbed round-trip propagator at the carrier
frequency for the PRFPMI shown in \fig{prifo_xfer} is
\begin{equation} \labe{kifo}
\begin{split}
K_{\text{IFO}} &(\Delta z_5, \Delta z_6) = e^{-i\, 2\, k\, \Delta
z_5}\, \\ &\times G_{56} \biggl[ T^{-\parallel}_6\, K_{616}(0)\,
T^{+\parallel}_6  \\ & + e^{-i\, 2\sqrt{2}\, k\, \Delta z_6}
R^{-\parallel}_6\, K_{626}(0)\, R^{-\perp}_6 \biggr] G_{65}\,
R^-_5
\end{split}
\end{equation}
where we have chosen to extract from \eqr{Rminus} and
\eqr{MIRbplus} the explicit phase offsets arising from nonzero
values of $\Delta z_5$ and $\Delta z_6$, respectively.

A close examination of \eqr{kifo} reveals that we can separate the
idealized servo-locking algorithm into three consecutive stages,
each with its own optimization process.
 \begin{enumerate}
 \item Follow the procedure described in \sct{FPIpseudolocker} to
 adjust individually the positions of $\mathcal{M}_1$ and
 $\mathcal{M}_2$ to maximize the round-trip enhancement for
 the lowest-loss carrier eigenmode of each FPI.
 \item Select an initial value for $\Delta z_6$, and then compute the carrier eigenvalue spectrum of
 $K_{\text{IFO}} (0, \Delta\, z_6)$ to find both the eigenvalue $\lambda_0$ with the largest magnitude
 and the corresponding
 lowest-loss eigenvector $E_0$. Set the Michelson input field $F^+_0 = G_{65}\,R^-_5\,E_0$
 in \eqr{PMI}, and then use \eqr{dz6} and the discussion in \sct{MIpseudolocker}
 to compute the new beamsplitter position $\Delta z_6^\prime$ that satisfies the dark port
 condition for $E_0$. Since $E_0$ is generally \emph{not } an eigenvector
 of $K_{\text{IFO}}(0, \Delta z_6^\prime)$, iterate this step to
 determine an eigenvector $E^\prime_0$ and a BS displacement $\Delta
 z_6^\prime$ such that successive values of the largest-magnitude
 eigenvalue $\left|\lambda^\prime_0\right|$ differ by no more than a suitably
 small convergence threshold (typically $10^{-12}$ in our
 simulations).
 \item  As in the case of the FPI discussed in \sct{FPIpseudolocker},
the eigenvector $E^\prime_{0}$ is also an eigenvector of the
matrix operator $K_{\text{IFO}}(\Delta z_5, \Delta z^\prime_6)$,
with the eigenvalue $\lambda_5 = \left|\lambda^\prime_{0}\right|
e^{i\, \left(\phi^\prime_0 - 2\, k\, \Delta z_5\right)}$. Hence,
if we choose
\begin{equation}
\Delta z_5 = \frac{\phi^\prime_{0}}{2\,k} ,
\end{equation}
then the net round-trip phase accumulated by $E^\prime_{0}$ will
be zero, and that mode will satisfy the resonance condition.
 \end{enumerate}
As in the case of the FPI, in our simulations we maintain a record
of the most recently chosen lowest-loss eigenvector, and in the
case of degeneracy we choose the eigenmode having the largest
value of the inner product with the previous eigenvector.

Suppose that we have followed this procedure and computed the
eigenvector $E^\prime_0$ of the matrix operator
$K_{\text{IFO}}(\phi^\prime_0/2\,k, \Delta z^\prime_6)$,
representing a maximally resonant field amplitude at reference
plane $z^<_5$ in \fig{prifo_xfer}. We can mode-match this field by
choosing an input field that optimally couples to $E_0$ through
the enhancement operator given by \eqr{Hp5}. If we solve the
matrix equation $E_0 = H^+_5(0)\, F^+_0$ for $F^+_0$, we obtain
 \begin{equation}
 F^+_0 = \left(\frac{1 - \left|\lambda_0\right|}{\lambda^*_0}\right)\,
 \left(T^-_5\right)^{-1} R^-_5\, E_0 ,
 \end{equation}
in agreement with \eqr{fpimm}.

 \subsection{Computation of Steady-State Fields\labs{IFOSS}}
In previous sections, we have been largely unconcerned with the
basis chosen to represent the operators which comprise the
transfer matrices of the optical elements of the
gravitational-wave interferometer. Instead, we have focused on the
details of specifying the operator algebra needed to describe the
physics of mirror perturbations. In this section, we intend to
define a computational algorithm which will allow us to determine
the self-consistent steady-state fields everywhere in a realistic
model of a thermally loaded interferometer. We begin by describing
a procedure to define an unperturbed transverse spatial basis that
can be used to represent all intracavity and extracavity
electromagnetic fields, and then we specify the initialization
process we have used to prepare the computational model for
perturbations. Finally, we present the simple iterative
steady-state solution algorithm we have used to obtain convergence
from the complete set of nonlinear coupled propagation equations
describing the PRFPMI intracavity fields.

In the Michelson interferometer, the beamsplitter couples two
optical systems (each consisting of a vacuum region followed by
either a mirror or a Fabry-Perot interferometer) that may have
significantly different mode-matching requirements relative to
some common reference plane. In principle, the determination of a
 set of eigenmodes that is common to both optical
systems is a tedious numerical exercise, particularly in the
context of the power-recycling scheme shown in
\fig{caltech_prifo}. However, in practice, our simulations do not
require the identification of an \emph{ab initio} set of common
eigenmodes. Rather, we seek a collection of self-consistent
unperturbed basis functions capable of accurately representing the
transverse spatial features of the laser field anywhere in the
gravitational-wave interferometer. As we show here, these basis
functions can be eigenmodes of the unperturbed parallel FPI
\emph{only}, propagated \emph{unidirectionally} throughout the
entire system.

Referring to \fig{caltech_prifo}, \fig{fpi_xfer}, and
\fig{mi_xfer}, we begin by ignoring all apertures everywhere in
the system, and choosing the length and mirror curvatures of the
unperturbed (e.g., infinite-aperture, spherical-mirror, and cold)
parallel FPI. These geometric configuration parameters define a
set of standing-wave eigenmodes with a particular spot size and a
wavefront radius of curvature at $\mathcal{M}_1$ that matches that
of $\mathcal{M}_1$ prior to thermal loading. \emph{We select these
eigenmodes as the spatial basis functions that will describe the
transverse laser field everywhere in the system.}

Referring to \fig{caltech_prifo} and \fig{prifo_xfer}, we position
another mirror at the location of the power recycling mirror (PRM)
$\mathcal{M}_5$ to couple optically the two Fabry-Perot
interferometers, and we propagate the lowest-loss (i.e.,
``fundamental'') basis function out of the parallel FPI through
the beamsplitter to the intracavity input reference plane $z^<_5$
of the PRM. We calculate the value of the curvature $1/R_F$ of the
extracted fundamental FPI eigenmode at $z^<_5$, and set the
unperturbed curvature of $\mathcal{M}_5$ to that value. We also
compute the spot radius $w_5$ of the fundamental basis function at
$z^<_5$ so that we can determine the perturbed curvature mismatch
matrices $C^\pm_5$ using the true curvature $1/R_5$ in \eqr{Cdef}
and \eqr{gammadef}.

Following the same unidirectional approach, we propagate the
fundamental basis function from the PRM intracavity output
reference plane $z^>_5$ to the intracavity reference planes
$z^>_2$ of $\mathcal{M}_2$ and then $z^>_4$ of $\mathcal{M}_4$ in
the perpendicular FPI. At each of these two reference planes, we
compute the curvature and spot radius of the corresponding laser
field, and then set these values as the appropriate unperturbed
parameters of these mirrors. As in the case of $\mathcal{M}_5$,
any discrepancies between the unperturbed curvatures of the
propagated basis functions and the true curvatures of the
perpendicular mirrors are then captured by the $C^\pm$
perturbation matrices defined by \eqr{Cdef}.

We complete our specification of the unperturbed geometry of the
interferometer by propagating a selected set of unperturbed
transverse spatial eigenmodes from the parallel FPI throughout the
PRFPMI. This procedure defines a (possibly incomplete) set of
spatial basis functions with which intracavity fields in the
PRFPMI can be specified. Furthermore, by propagating the
intracavity basis functions out through both the dark port and
$\mathcal{M}_5$, we can use the same unperturbed basis for the
extracavity fields $F^+$, $E^+$, and $E^-$ shown in
\fig{prifo_xfer}. In general, the input field $F^+$ will not be
described solely by the outward-propagated fundamental eigenmode
of the parallel FPI; instead, it will be represented by a linear
combination of some subset of the modes available in the
unperturbed basis.

The spherical-mirror, infinite-aperture, cold approximation of the
unperturbed gravita\-tional-wave interferometer lends itself to a
convenient representation of intracavity fields by
power-orthogonal Hermite-Gauss modes. However, even when the
mirrors remain spherical and thermal lensing is ignored, this
basis cannot represent all possible perturbed fields with
arbitrarily high accuracy under all circumstances. For example, in
cases where either the recycling cavity is geometrically unstable
or a thermal lens is so strong that the sign of the curvature of a
wavefront propagating through the substrate changes, then we
expect that the number of stable unperturbed basis functions
needed to represent that field will become prohibitively large.
Furthermore, if either finite apertures or non-spherical mirrors
are used to define the unperturbed basis numerically, it is
possible that residual diffraction of a non-Hermite-Gauss
fundamental basis function from $\mathcal{M}_5$ (followed by
propagation to $\mathcal{M}_6$) will generate a field profile that
has a mean spot radius that differs from the value computed using
the field profile obtained by propagating the lowest-loss parallel
eigenmode from $\mathcal{M}_1$ to $\mathcal{M}_6$. In this case, a
biorthogonal basis can be chosen, but special care must be taken
to ensure that the eigenmode expansions converge.\cite{kos97} When
a self-consistent basis can be chosen, and the corresponding
perturbation expansion of the physical observables under
simulation converges, the matrix representation of the problem can
allow computations which are orders of magnitude faster than those
performed by FFT codes.

Now that we have defined a self-consistent set of spatial basis
functions at every reference plane in the LIGO PRFPMI, we can
introduce the perturbations by setting the apertures and
curvatures of the mirrors and beamsplitter to their true values,
and then computing the matrices that characterize the effects of
aperture diffraction, wavefront-mirror curvature mismatch, and
thermal focusing on laser fields expressed as linear combinations
of the \emph{unperturbed} basis functions. Initially, we solve
\eqr{E5qifo} in the limit $F^+ \rightarrow 0$ to obtain the
perturbed fields of the cold interferometer, and then we choose a
relatively low but finite input power (e.g., 0.1~W) for the input
field. The steady-state intracavity fields under this small
thermal load can be found using a straightforward, self-consistent
procedure:

\begin{enumerate}
\item Using the current intracavity fields, compute the powers absorbed
in the mirror and beamsplitter substrates and coatings, and update
the substrate thermal OPD matrix elements given by \eqr{S} and
\eqr{Sbs}, which depend nonlinearly on these powers.
\item Following the procedure outlined in \sct{IFOpseudolocker},
adjust the micropositions of $\mathcal{M}_1$, $\mathcal{M}_2$,
$\mathcal{M}_6$, and $\mathcal{M}_5$ to achieve optimum resonant
phase conditions for the lowest-order carrier mode in the each of
the three optical cavities.
\item Recompute the intracavity fields using \eqr{E5qifo}.
\item Repeat the previous steps until the recycled power $\half \sum_q \left|E^-_{5 q}\right|^2$ has stabilized.
\end{enumerate}

Note that we do \emph{not} propagate the field from one reference
plane to another throughout the interferometer. In this case, even
when using the simulated phase control system, the fields tend to
converge after a number of iterations implied by the photon
storage lifetimes of the arm cavities. Furthermore, we do not use
more complicated nonlinear gradient-search methods to determine
the intracavity fields because the number of variables (field
basis function coefficients) is extremely large, and the variables
at any single location in the interferometer is linked to those at
all other locations by nontrivial round-trip propagators. Instead,
the solution procedure we have detailed above can provide
convergence to a few parts in $10^4$ in only a few iterations.

The steady-state fields arising from higher input powers can be
found by initializing the convergence process with scaled field
amplitudes computed at a lower power. For example, if a stable
solution has been found at 2~W total input power, then the
\emph{same} collection of intracavity fields, scaled by a factor
of two, would be reasonable initial values for an input power of
8~W. Similarly, if the input power is held constant, but some
other parameter of the PRFPMI configuration (such as the PRM
curvature) is varied, the intracavity fields can be re-used from
one parameter value to the next. In this way, a set of solutions
for a variety of different interferometer parameter sets can be
constructed in a few minutes using a typical desktop computer.

An object-oriented numerical computer model of the initial LIGO
interferometer based on these algorithms has been implemented
using the class mechanism of the MATLAB programming
language.\cite{matlab} The resulting collection of MATLAB source
code files, hereafter referred to as ``Melody,'' is available
publicly for examination and execution.\cite{mel01}

\subsection{Comparison with a Fast Fourier Transform Model\labs{fft}}
\begin{table}
\caption{High-reflection ($\mathit{hr}$) and anti-reflection
($\mathit{ar}$) loss parameters for the initial LIGO optical
elements used in our simulations. Here $a_c$ is the optical power
absorption that contributes to the substrate thermal loss, and
$s_c$ is the scattering loss.\labt{coatings}}
\begin{ruledtabular}
\begin{tabular}{ccccrc}
& Parameter & & \multicolumn{3}{c}{Loss (ppm)}\\
 \hline
 & $a_{hr}$ & & & 1.0 &\\
 & $a_{ar}$ & & & 1.0 &\\
 & $s_{hr}$ & & & 90.0 &\\
 & $s_{ar}$ & & & 600.0 &\\
\end{tabular}
\end{ruledtabular}
\end{table}

\begin{table}
\caption{Physical parameters for the initial LIGO optical elements
used in our simulations. Each substrate is fused silica, with
parameters given by \tbl{silica} and a radius $a = 25$~cm. Here
the total loss $l_{hr} = a_{hr} + s_{hr}$, where $a_{hr}$ and
$s_{hr}$ are the power absorption and scattering losses in the
high-reflection coating, respectively.\labt{mirrors}}
\begin{ruledtabular}
\begin{tabular}{crccc}
Mirror & \multicolumn{1}{c}{$R_M$~(m)} & $h$~(m) & $r^2$ & $t^2$\\
 \hline
$\mathcal{M}_1$, $\mathcal{M}_2$ & 14 571.0 & 0.10 & 0.970000 & $1
- r^2 - l_{hr}$\\
 $\mathcal{M}_3$, $\mathcal{M}_4$ & 7 400.0 & 0.10 & $1 - t^2 - l_{hr}$ & $1.5 \times 10^{-5}$ \\
 $\mathcal{M}_5$ & 9 999.8 & 0.10 & 0.985020 & $1 - r^2 - l_{hr}$\\
 $\mathcal{M}_6$ & \multicolumn{1}{c}{$\infty$} & 0.04 & 0.499975 & $1 - r^2 - l_{hr}$\\
\end{tabular}
\end{ruledtabular}
\end{table}

In certain limited cases, the predictions made by Melody can be
compared with those of a Fortran implementation of a Fast Fourier
Transform (FFT) model of the initial LIGO
interferometer.\cite{fft98} The FFT model was designed to
investigate geometric requirements and tolerances for initial LIGO
optical components, and allows either a measured or simulated
static OPD phase map to be included in each reflective surface. By
contrast, the Melody solution method described in \sct{IFOSS}
allows the evolving fields everywhere in the interferometer to
alter the local phase maps using nonlinear thermal lensing, so
that both the field and mirror are distorted by their reciprocal
interaction. For the FFT the paraxial approximation is used in
order to calculate the propagators between mirrors, as matrix
operators in the spatial frequency domain. The lengths are
optimized in order to achieve a stationary locked configuration,
and the power stored inside the arms and the recycling cavity is
evaluated. The final fields obtained may be analyzed for modal
composition. In contrast, the Melody program starts with a fixed
number of paraxial modes. The matrix elements coupling the modes
are then analytically calculated using the algorithms described in
\sct{mirrors} and \sct{beamsplitter}, with the thermal effects
perturbatively evaluated at every iteration until stationarity is
achieved.

While the basic optics and wave-equation assumptions are the same,
the two programs have markedly different implementations. In the
case of the FFT code, the carrier and the sidebands are simulated
by two separate, consecutive runs so that the cavity lengths are
optimized for the carrier, and the dimensions of the recycling
cavity are optimized for the sidebands. As discussed previously,
the idealized Melody control system sets the cavity lengths by
ensuring a round trip phase of zero for the lowest-loss eigenmode.
The FFT solutions scale linearly with initial field power, but in
Melody, the power is shared by carrier and sidebands; the carrier
and sideband fields are separately calculated, and their combined
power at each iteration affects the thermal lens distortion
nonlinearly. The FFT features fine-grained meshing of fixed phase
map information, being intended primarily to study arbitrary
mirror aberrations. On the other hand, Melody is intended to much
more rapidly study coarser-grained effects with specific emphasis
on the non-linear effects of thermal focusing. Significantly, the
computational hardware requirements for the codes is dramatically
different: a typical FFT run requires 30--60 minutes on a
supercomputer, while a typical Melody computation requires only a
few seconds to stabilize.

\epsf{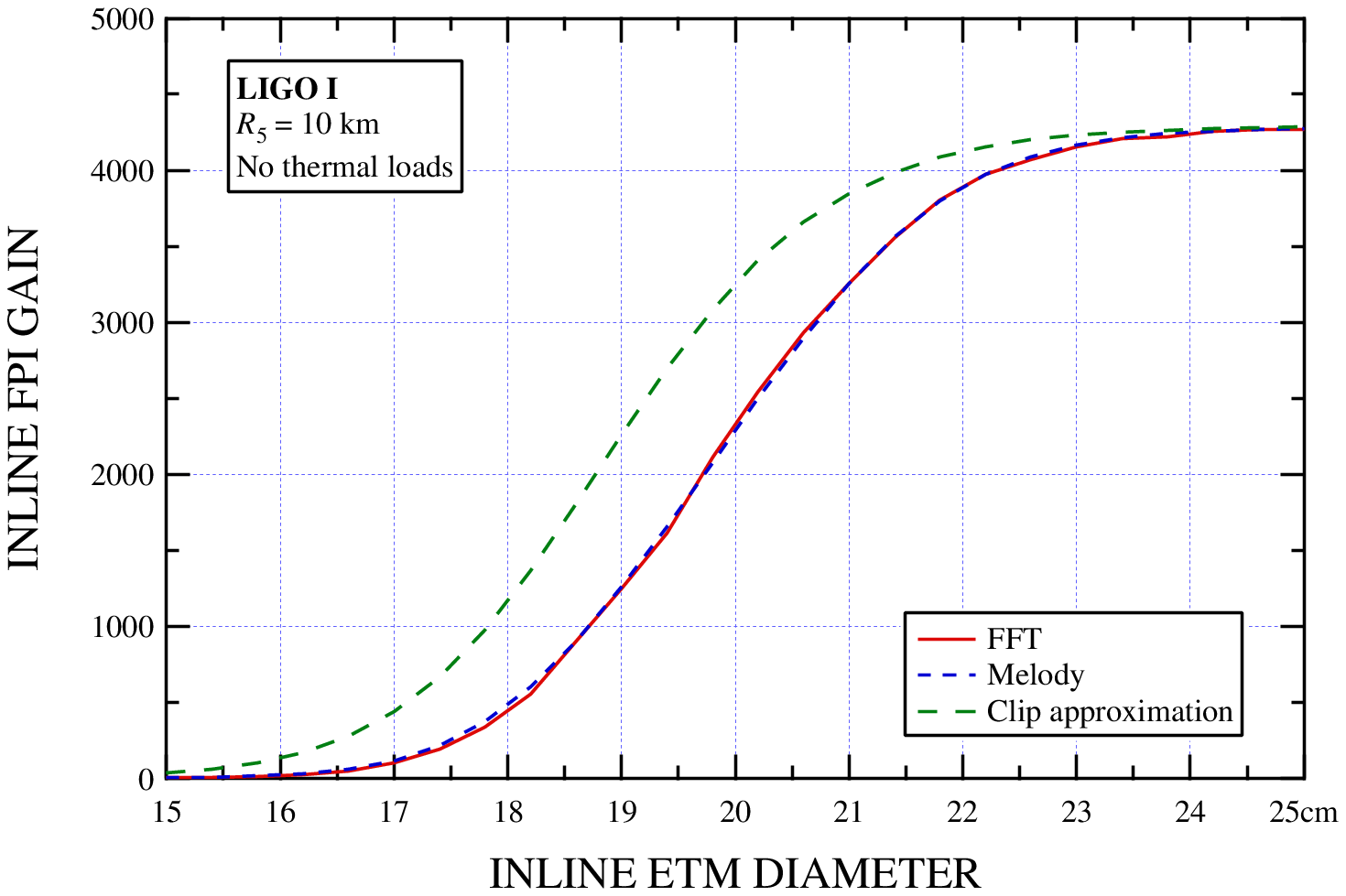}{A comparison between the predictions of
computer code based on our model (Melody) and those of the FFT
model in the case of simple aperturing. The power enhancement (or
``gain'') at the parallel/inline FPI reference plane $z^>_1$ in
\fig{fpi_xfer} 
is plotted as a function of the aperture of $\mathcal{M}_3$. The
``clip approximation'' represents the result expected by assuming
that the loss arises from simple single-pass aperture clipping at
$\mathcal{M}_3$.}

The parameters that we used in our simulations are shown in
\tbl{coatings} and \tbl{mirrors}. The length of each arm cavity
was chosen to be 3999.01~m, and the RF modulation frequency
$\omega_1/2\pi \equiv 24.493$~MHz (with a modulation depth of
0.44) was adjusted slightly in the Melody computations to
guarantee an antiresonance in the arm cavity to within machine
precision. The current design value of the initial LIGO common
length (the average of the distances between the power recycling
mirror and the input test masses) is 9.188~m, and the differential
length (or "Schnupp") asymmetry (the difference between the
distances between the parallel and perpendicular ITM and the PRM)
is $L_S = 0.15$~m. The distance between the beamsplitter and the
PRM has the value 6.253~m, and all simulations were performed
using the Nd:YAG laser wavelength $\lambda_0 = 1.0642$~$\mu$m. As
a first comparison, both codes were used to simulate the
unperturbed initial LIGO interferometer, with perfect mode
matching, infinite-diameter optics, and no thermal distortion. The
agreement between computations of reflected and transmitted power
with analytical calculations are within 0.2\% for both the FFT
code and Melody.

\epsf{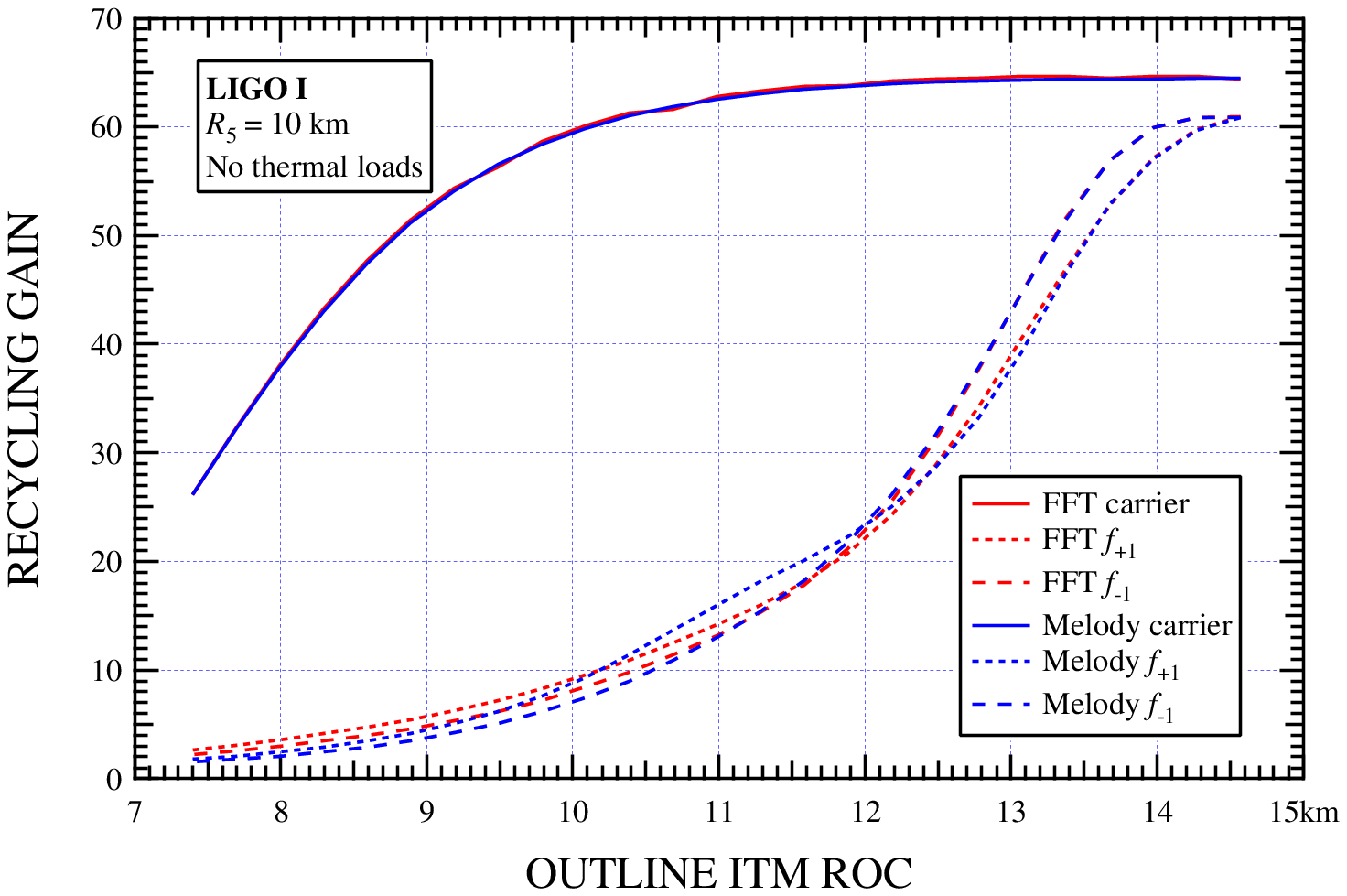}{A comparison between the predictions of
Melody and those of the FFT model in the case of unstable
curvature reduction. The gain at the recycling mirror reference
plane $\mathcal{A}^+_5$ in \fig{prifo_xfer} 
is plotted as a function of the curvature of $\mathcal{M}_2$. Note
that --- in the absence of thermal loading
--- the recycling cavity becomes unstable for the resonant sidebands at a
radius of curvature of 14480~m.}

In the absence of thermal loading, we have studied the dependence
of the parallel (inline) arm cavity power buildup on the diameter
of the ETM $\mathcal{M}_3$. A comparison between the predictions
of Melody and the FFT code tests the suitability of the former's
basis-function-expansion approach to calculating interferometer
power levels, particularly when the cavity losses are large enough
to require the inclusion of a large number of modes. In
\fig{melody_fft_aperture}, we plot the ratio of the total
circulating optical power at reference plane $\mathcal{A}^+_1$ in
\fig{fpi_xfer} to the total carrier input power $\half |F^+_0|^2$.
When the 28 lowest-loss Hermite-Gauss modes (corresponding to $m +
n \le 6$) of the parallel FPI are used, the agreement is excellent
even at an ETM diameter of only 15~cm, which imposes a round-trip
diffractive loss of approximately 2000~ppm. For comparison, we
have included a trace displaying the stored power in the parallel
arm expected when the intracavity loss is due to simple
single-pass aperture clipping. Note that this naive assumption
underestimates the true diffractive loss by a factor of about 2.14
for this beam spot size at $\mathcal{M}_3$.

In the thermally loaded simulations that are described in
\sct{ifo_sim}, the increase in thermal focusing power of the ITM
substrates improves the geometric stability of the power recycling
cavity by increasing the apparent ITM radii of curvatures.
However, if we reduce the radius of curvature (ROC) of either ITM
below the design value of 14.571~km, the recycling cavity becomes
unstable, resulting in a large reduction in stored power. Previous
studies of this phenomenon using the FFT code have shown that ---
for small changes in ITM curvature --- only the sidebands suffer
this power degradation; the carrier resonance condition in the
corresponding arm cavity serves to stabilize the recycling cavity
mode by attenuating the anti-resonant higher-order modes.
Therefore, since our model uses stable resonator eigenmodes as the
spatial basis functions for the field expansions, this is a much
more strenuous test than the aperture reduction example, requiring
a substantially greater number of transverse modes. In
\fig{melody_fft_roc} we show the dependence of the power recycling
cavity gain on the perpendicular (outline) ITM radius of
curvature. The Melody results for the carrier agree essentially
perfectly with those of the FFT code even when we employ only the
28 lowest-loss Hermite-Gauss modes of the parallel FPI. However,
to obtain reasonably small discrepancies between the predictions
of the sideband behavior in the highly unstable regime, we must
include all basis functions satisfying $m + n \le 20$, or 231
modes.



\subsection{Initial LIGO Optical Performance Simulations\labs{ifo_sim}}
We now turn to simulations of the basic initial LIGO
interferometer in the presence of thermal loading of the optics.
Because of the high power levels maintained in the interferometer
to suppress shot noise, the absorption of power in the ITM
substrates and coatings and the subsequent thermal focusing must
be incorporated into the interferometer design. Using Melody, we
explore in detail the effects of this focusing on the optical
performance of the interferometer. Since the recycling cavity is
operating in a region of geometric stability under full thermal
load, we have run our simulations using the 28 lowest-loss
Hermite-Gauss modes (corresponding to $m + n \le 6$) of the
parallel FPI.

If the radius of curvature of the power recycling mirror
$\mathcal{M}_5$ is chosen to be the cold-cavity mode-matched value
shown in \tbl{mirrors}, the thermal distortions caused by the
substrates in the recycling cavity will destroy this mode-matching
and significantly reduce the available recycling gain for the
sidebands. In \fig{ligo_i_recycled_sb_power_65} and
\fig{ligo_i_output_sb_power_65}, we plot the sideband recycled
powers $|F_{5,\pm 1}|^2$ given by \eqr{E5qifo} as a function of
the PRM ROC $R_5$ for the initial LIGO TEM$_{00}$ operating input
laser power of 6.5~W, with 5.9~W delivered by the carrier when the
sideband modulation depth is 0.44. The presence of the arm
cavities significantly reduces the sensitivity of the carrier to
these perturbations, resulting in an approximately constant
recycling gain of 30 when the parameters specified in
\tbl{silica}, \tbl{mirrors}, and \tbl{coatings} are used. Note
that the optimum mode-matching of the parallel and perpendicular
segments of the recycling cavity occurs when the PRM ROC has
increased to approximately 15~km.

One of the most striking details of
\fig{ligo_i_recycled_sb_power_65} and
\fig{ligo_i_output_sb_power_65} is the significant difference in
the recycling gain of the upper ($+\Delta \omega_1$) and lower
($-\Delta \omega_1$) sideband fields. Such a sideband imbalance
can lead to significant gravitational-wave detection noise
signatures that are qualitatively absent when the balance is
exact. Since the ratio $\Delta \omega_1/\omega_0$ is immeasurably
small for RF sidebands and (by design) there is a very high degree
of geometrical symmetry between the Michelson branches of the
interferometer, we naively expect the sideband fields to interact
identically with geometrical distortions as long as these
distortions are frequency-independent at the RF scale. In
principle, this symmetry is broken only by the macroscopic value
of the Schnupp asymmetry $L_S \equiv L_{51} - L_{52}$, where $L_{5
j}$ is the optical path length separating the $\mathcal{M}_5$ and
$\mathcal{M}_j$. For gravitational-wave interferometers, $L_S$ is
much smaller than the Rayleigh length, so both sidebands should
interact nearly identically with either arm, except for a phase
accumulation proportional to the distortion-independent factor
$\Delta \omega_1 L_S$. Using both Melody and the FFT code, we have
determined that our simulations strictly show excellent sideband
balance if either $L_S \rightarrow 0$ \emph{or} the net branch
distortions are identical. However, with Melody we have determined
that the branch distortions \emph{cannot} be identical when the
thermal lens in the beamsplitter substrate is significant; this
source of additional distortion in the parallel arm requires a
significant displacement of the beamsplitter to maintain the dark
port condition, and is therefore the primary cause of the
imbalance shown in \fig{ligo_i_recycled_sb_power_65} and
\fig{ligo_i_output_sb_power_65}. In fact, as seen in
\fig{melody_fft_roc}, the FFT code reproduces this behavior when
the field-mirror curvature mismatch at the perpendicular ITM
becomes large, causing a significant difference between the net
distortions of the two arms. Further studies using Melody have
shown that it is possible to improve the sideband balance by
supplementing the procedure outlined in \sct{IFOpseudolocker} with
an additional step that readjusts the positions of both
$\mathcal{M}_5$ and $\mathcal{M}_6$, at the expense of increasing
the TEM$_{00}$ carrier output power.

\epsf{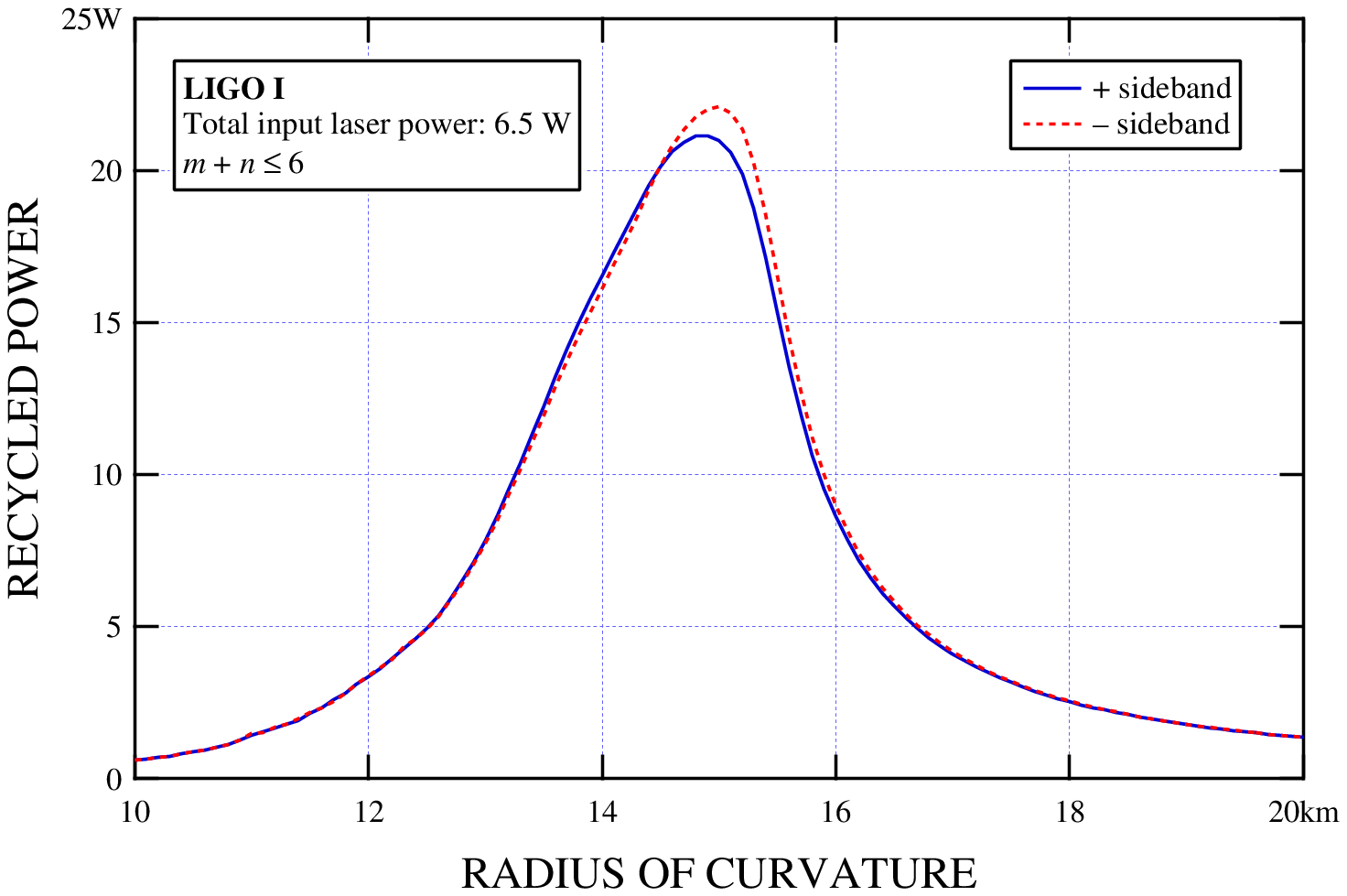}{Initial LIGO recycled powers as
a function of the PRM radius of curvature. The total TEM$_{00}$
input laser power is 6.5~W, with 5.9~W delivered by the carrier.}

\epsf{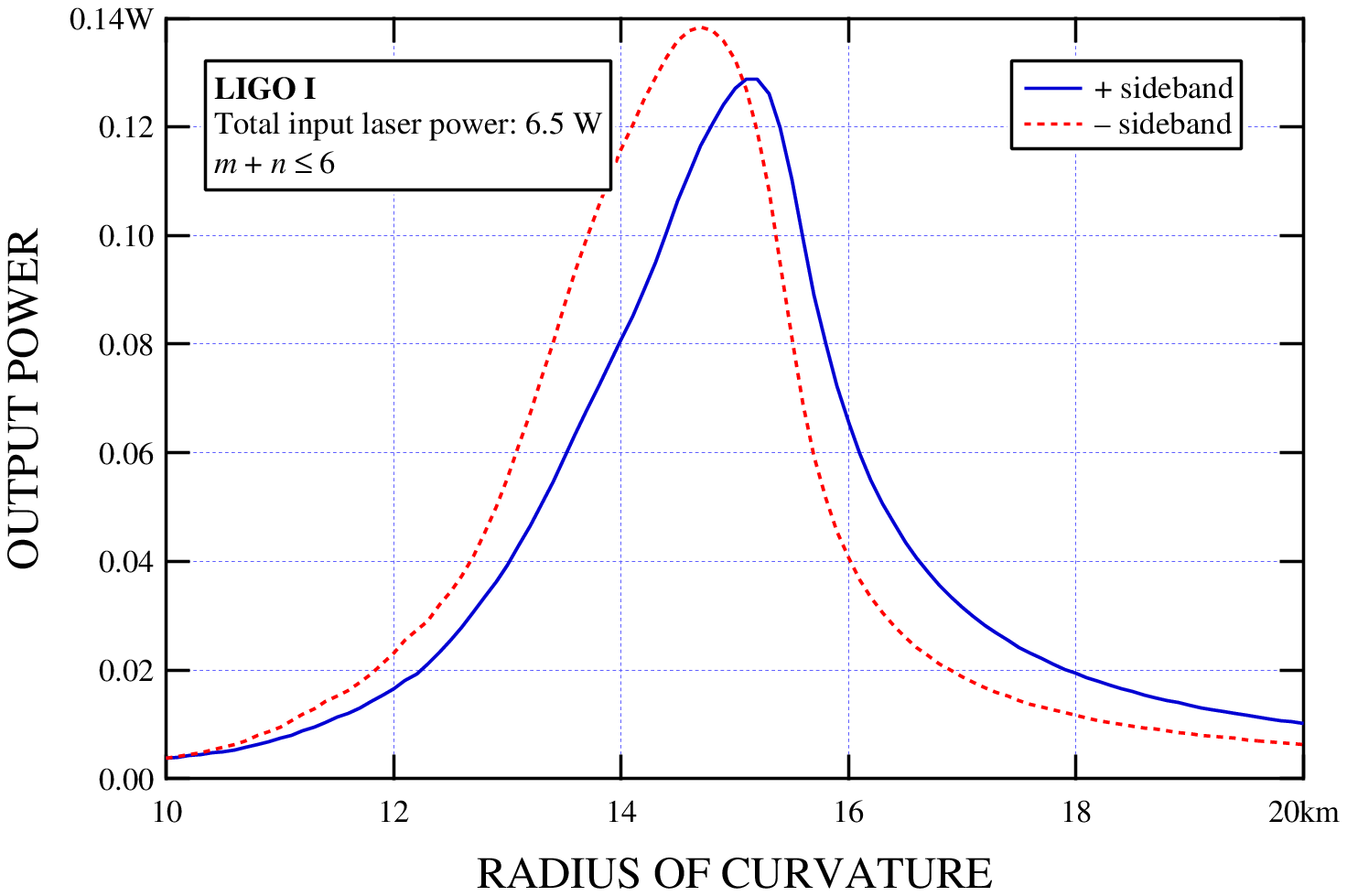}{Initial LIGO dark port powers as
a function of the PRM radius of curvature. The total TEM$_{00}$
input laser power is 6.5~W, with 5.9~W delivered by the carrier.}


For the purposes of our discussion here, we have chosen a value of
the PRM ROC that minimizes both the fundamental carrier power and
the sideband imbalance at the output port. This strategy produces
the optimum initial LIGO PRM radius of curvature as a function of
total TEM$_{00}$ input laser power shown in \fig{opt_prm_roc}.
Again, we have assumed a modulation depth of 0.44, which pushes
about 10\% of the total available laser power into the sidebands.
The corresponding optimized initial LIGO recycled and output
sideband powers are shown in \fig{opt_recycled_sb_power} and
\fig{opt_output_sb_power}, respectively. Note that at each value
of the input power in these two plots, the corresponding PRM ROC
shown in \fig{opt_prm_roc} has been used. Although the residual
imbalance of the recycled sideband fields remains at the 10\%
level at an input power of 10~W, the imbalance at the
antisymmetric port is negligible.

As an example of the operating characteristics of a fixed
interferometer design point, \fig{ligo_i_good_pr} and
\fig{ligo_i_good_dp} give the recycled and output powers at the
carrier and sideband modulation frequencies for an interferometer
optimized for a total laser input power of 6.5~W. The optimum PRM
ROC is 15.075~km, and, as shown in \fig{ligo_i_good_pr}, the
recycling gain for the carrier is about 30.
Figure~\ref{f:ligo_i_good_dp} predicts that, at 6.5~W, the
sidebands are approximately balanced, with about 130~mW in each
field, while the total carrier power emerging from the dark port
is about 180~$\mu$W. Approximately 90\% of this carrier power is
stored in higher-order modes.

The behavior of the idealized PRFPMI phase-control system is
illustrated in \fig{ligo_i_good_dz}. The PRM ($\mathcal{M}_5$) is
initially pulled in toward the beamsplitter to compensate for the
longitudinal phase introduced by the mismatch between the
curvatures of the field emerging from the parallel FPI and the PRM
itself. However, as the strengths of the thermal lenses in the ITM
substrates increase, the mirror is pushed away from the
beamsplitter to maintain the carrier resonance. Similarly, as the
astigmatic thermal lens develops in the beamsplitter substrate,
the beamsplitter moves slightly toward the upper left in
\fig{caltech_prifo}. Indeed, as we see in \fig{ligo_i_good_fl},
even in the $y$ plane the effective focal length of the
beamsplitter substrate is six times larger than that of either ITM
substrate; nevertheless, the resulting asymmetry between the two
Michelson arms is numerically detected and compensated.

\epsf{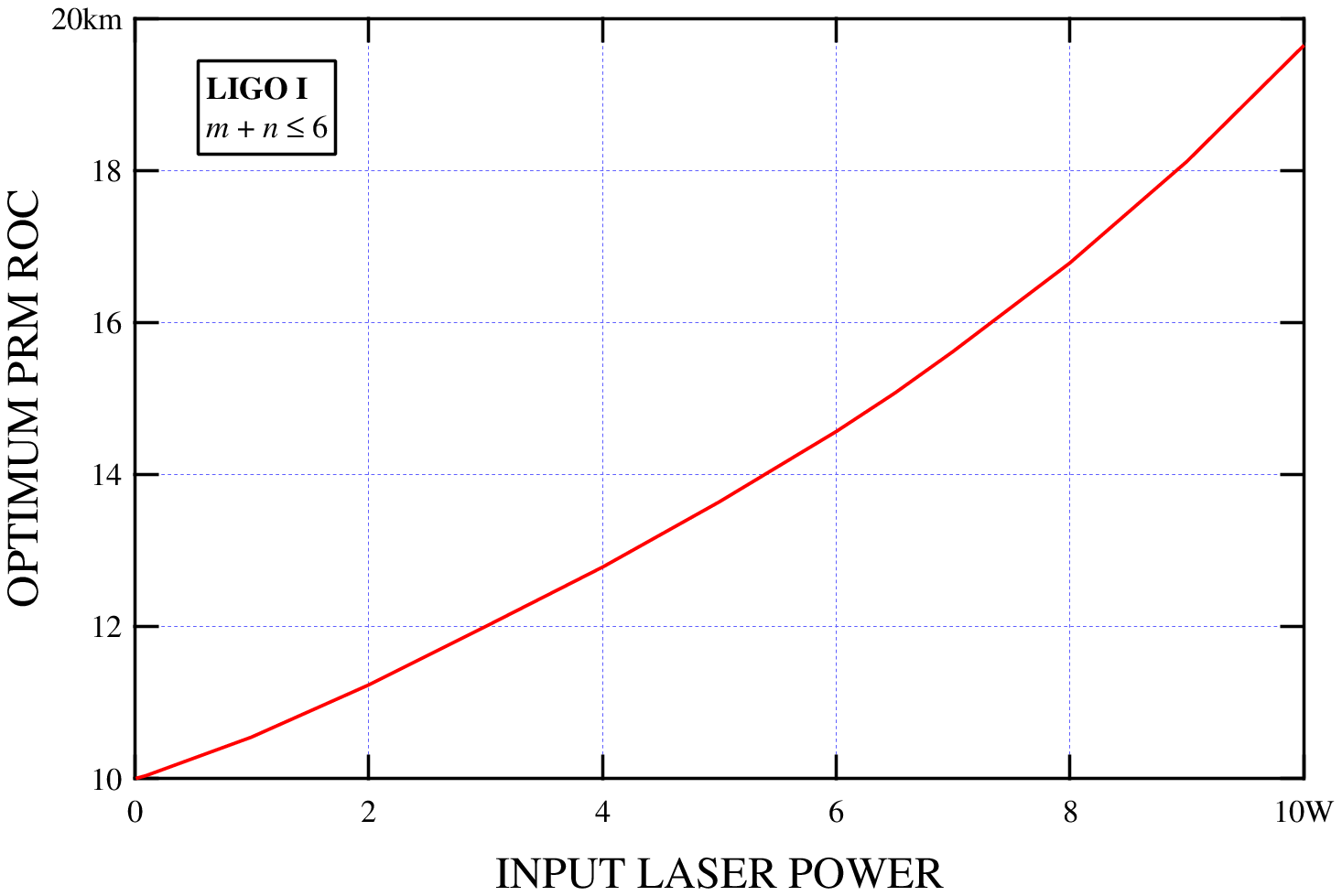}{The optimum initial LIGO PRM radius of
curvature as a function of total TEM$_{00}$ input laser power.}

\epsf{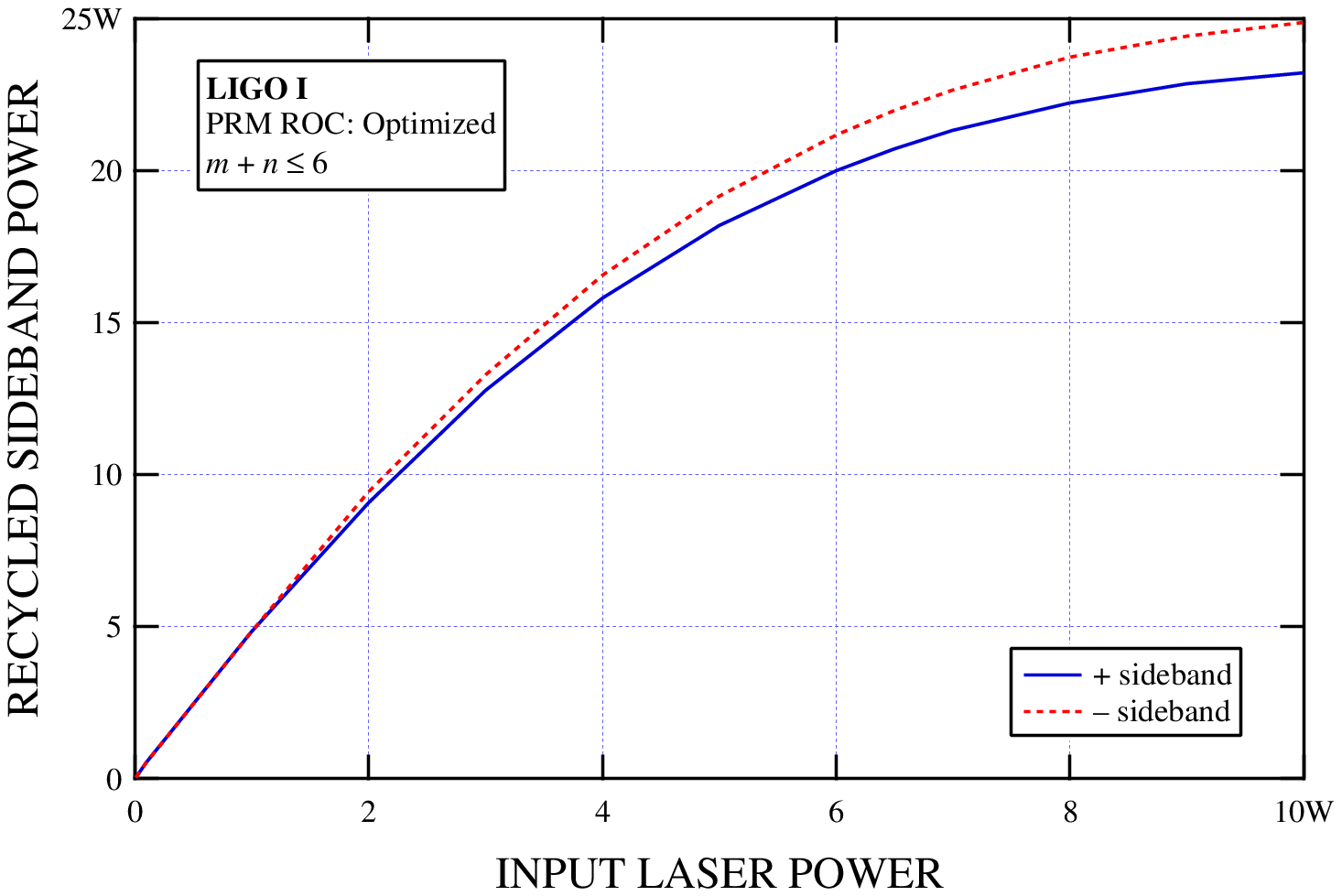}{Optimized initial LIGO recycled
sideband powers as a function of total TEM$_{00}$ input laser
power. At each value of the input power, the PRM radius of
curvature shown in \fig{opt_prm_roc} has been used.}

\epsf{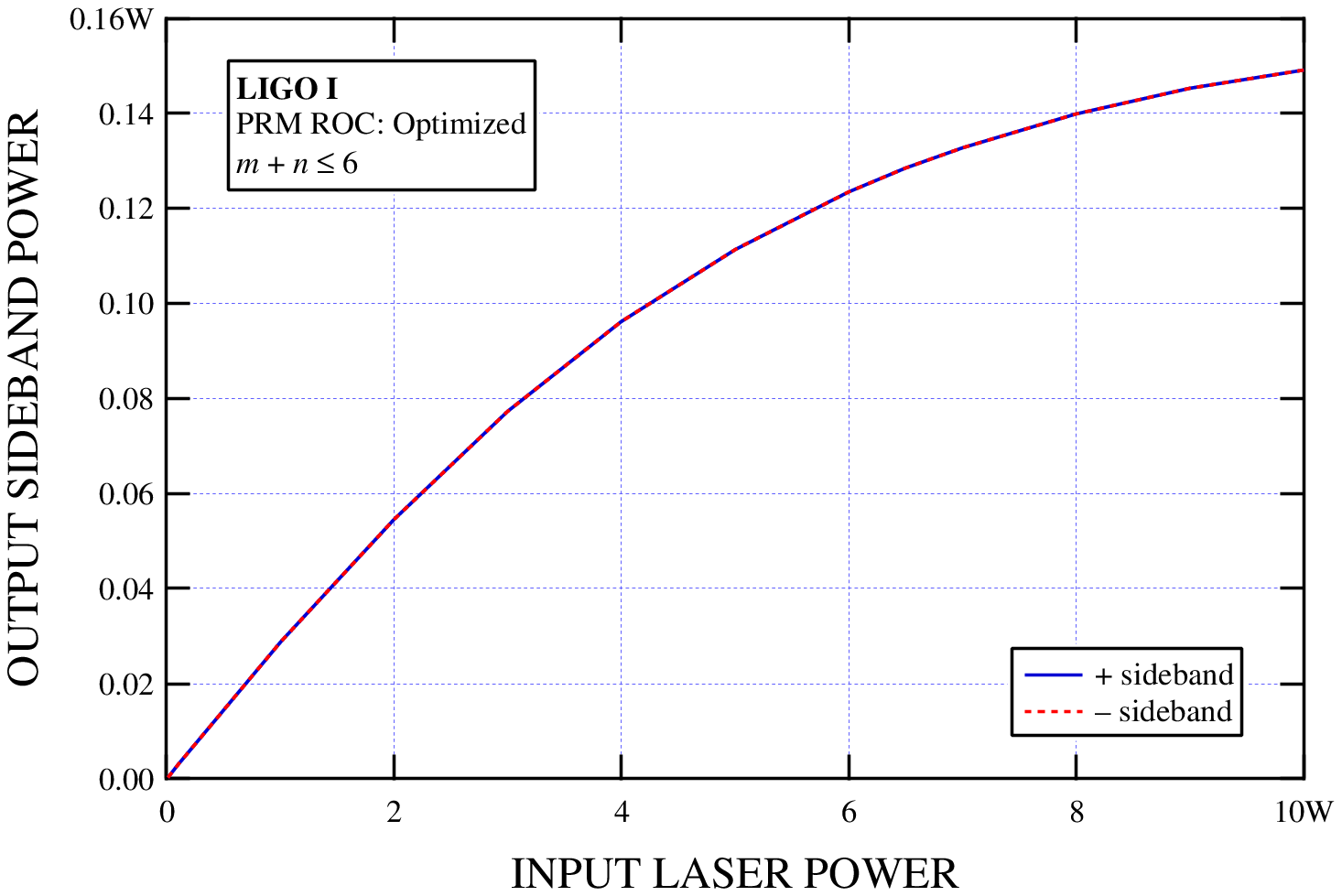}{Optimized initial LIGO output sideband
powers as a function of total TEM$_{00}$ input laser power. At
each value of the input power, the PRM radius of curvature shown
in \fig{opt_prm_roc} has been used.}

\epsf{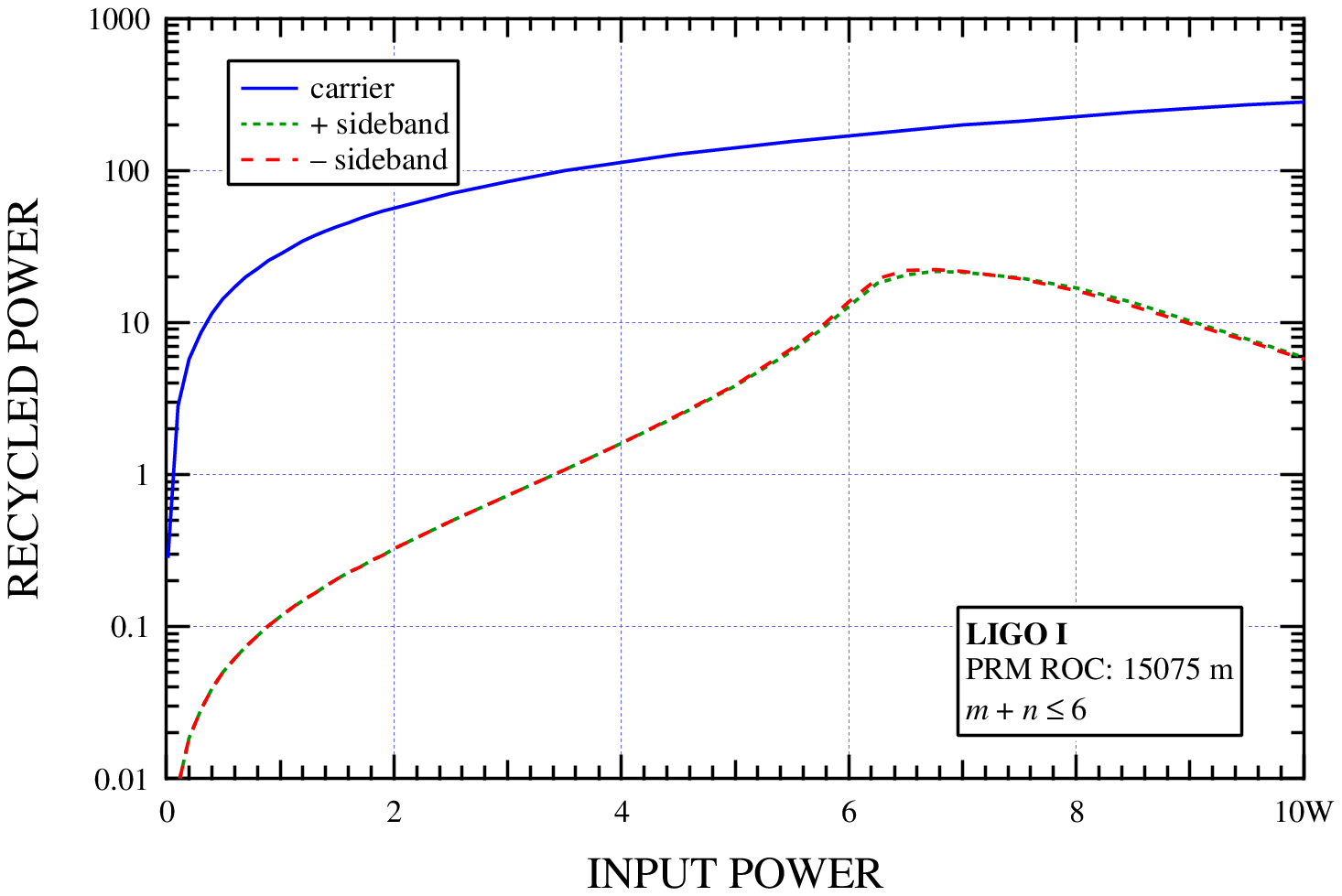}{Initial LIGO recycled powers as a function
of total TEM$_{00}$ input laser power. The PRM radius of curvature
is 15075~m, the optimized value for an input power of 6.5~W.}

\epsf{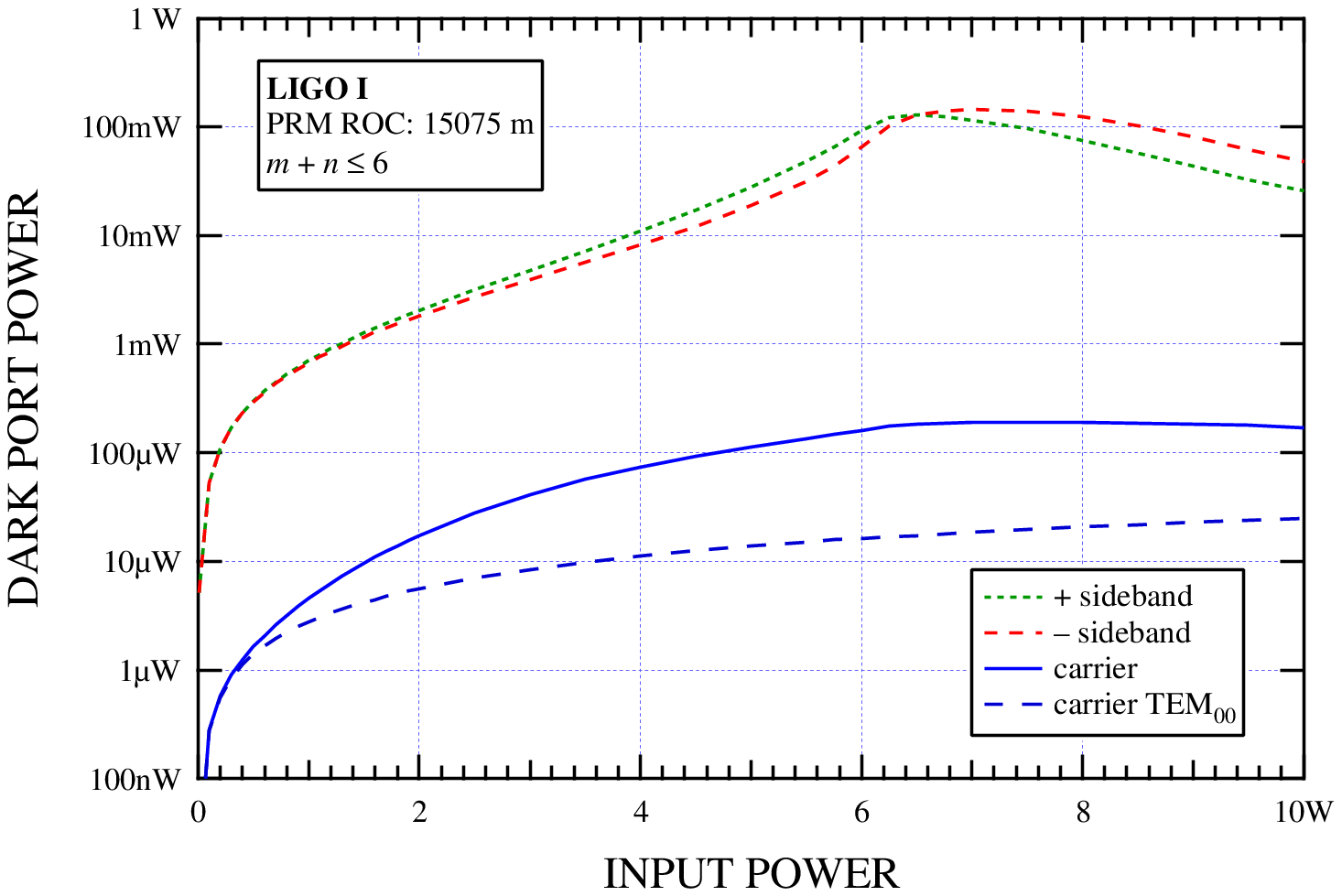}{Initial LIGO dark port powers as a function
of total TEM$_{00}$ input laser power.}

\epsf{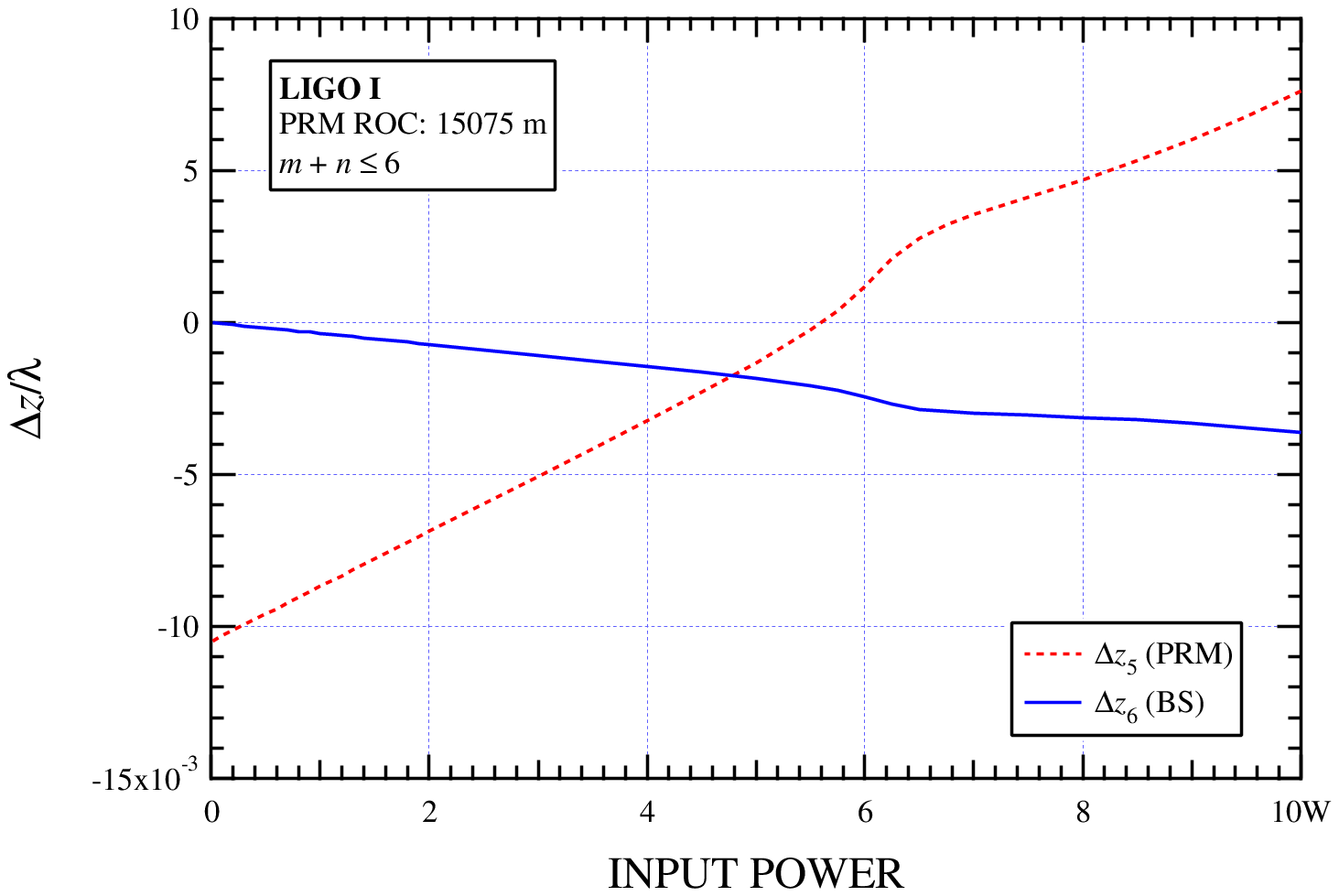}{Initial LIGO PRM and BS microdisplacements
as a function of total TEM$_{00}$ input laser power.}

\epsf{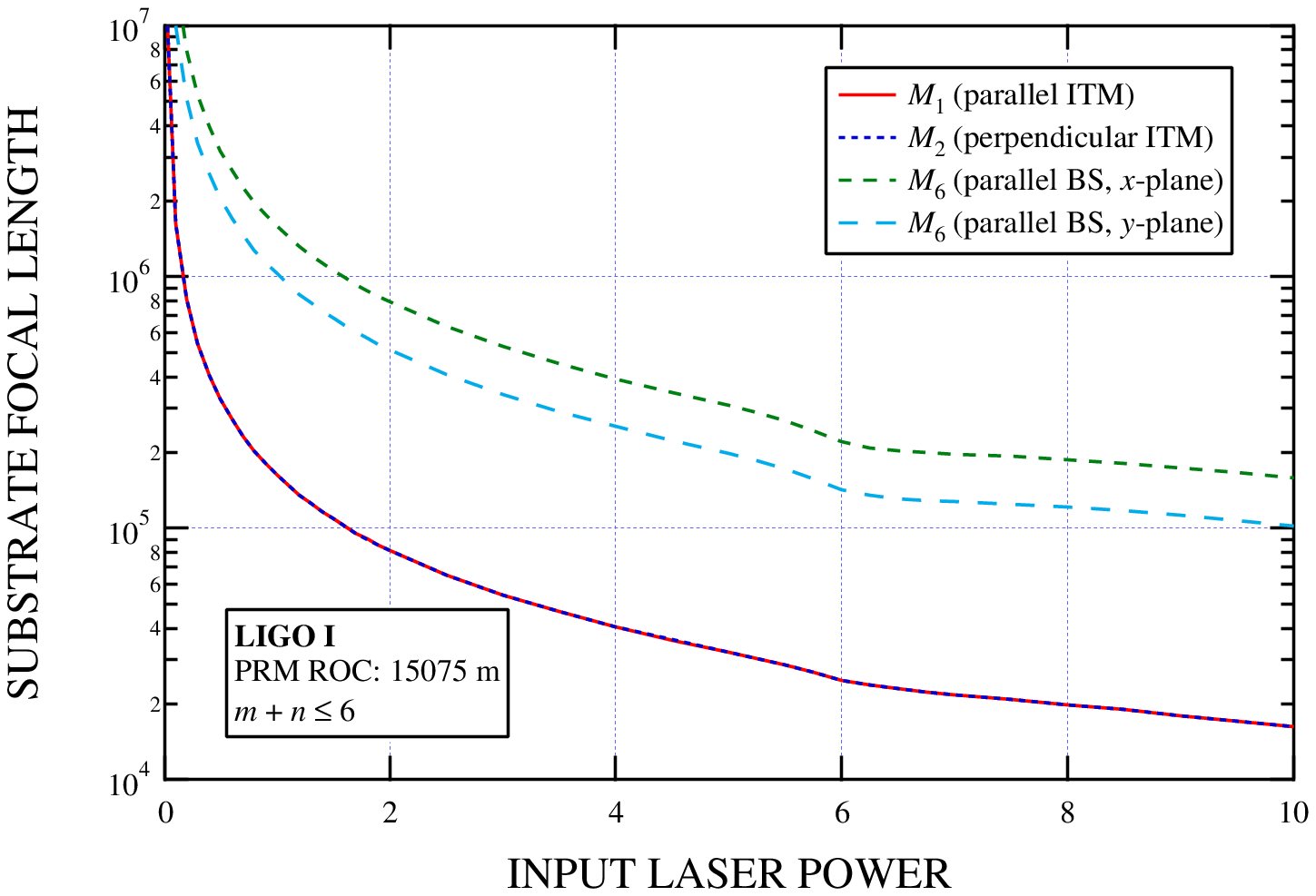}{Initial LIGO effective (i.e., quadratic)
thermal focal lengths as a function of total TEM$_{00}$ input
laser power.}



 \section{Conclusion}

We have developed highly detailed steady-state analytical and
numerical models of the physical phenomena which limit the
intracavity power enhancement of recycled Michelson Fabry-Perot
laser gravitational-wave detectors. We have included in our
operator-based formalism the effects of nonlinear thermal lensing
due to power absorption in both the substrates and coatings of the
mirrors and beamsplitter, the effects of any mismatch between the
curvatures of the laser wavefront and the mirror surface, and the
diffraction by an aperture at each instance of reflection and
transmission. We have taken great care to preserve numerically the
nearly ideal longitudinal phase resonance conditions that would
otherwise be provided by an external servo-locking control system.
Once the mirrors and beamsplitters have been analyzed and their
transfer matrices constructed, more complicated optical components
and structures (such as Fabry-Perot and Michelson interferometers,
and the LIGO detector) can be modeled using simple matrix
composition.

Our formalism relies on the validity of an expansion of the
intracavity electromagnetic field at any reference plane in a
necessarily incomplete set of possibly biorthogonal basis
functions that are derived from eigenmodes of one of the
Fabry-Perot arm cavities. In two specific cases, we have checked
the results produced by Melody --- a fast MATLAB implementation of
our model --- against those generated by an FFT program used to
model optical imperfections in LIGO. We have found that Melody can
accurately predict the behavior of the carrier even in highly
perturbed cold resonators using a modest number of spatial basis
functions in orders of magnitude less time than is required by the
FFT code. The same is true for the sideband fields stored in the
power recycling cavity, except in the case where the recycling
cavity is significantly geometrically unstable. Even in this
situation (which does not arise when the interferometer is
thermally loaded), accuracy can be improved substantially by
including more modes in the simulation.

We have discovered and described in broad terms the conditions
under which the power stored in the two sideband fields can become
unbalanced, potentially affecting the noise sensitivity of the
gravitational-wave interferometer in the detection band. At this
point, we believe that an additional post-process step can be
introduced into our simulated phase-locking algorithm that will
allow a reduction of the sideband imbalance at the expense of a
previously optimized optical property of the carrier field.

 In the future, we will include the effects of
thermoelastic surface deformation in our model, and apply it to
proposed configurations of the advanced LIGO interferometer. Other
systematic perturbations and imperfections (e.g., mirror tilt and
substrate inhomogeneities) can be included easily by incorporating
the appropriate operators into the transfer matrices describing
reflection and transmission for the mirrors and beamsplitter. In
addition, we intend to develop algorithms for the simulation of
the optical response of an interferometer to gravitational
radiation, with the intention of estimating the detection
sensitivity of advanced LIGO under full thermal load. We
anticipate that the performance of the Melody program will not be
significantly affected by these modifications.

\acknowledgments

 We thank Ryan Lawrence for providing us with the
results of his numerical computations of the temperature
distributions in the LIGO beamsplitter, and we are grateful to the
Center for Advanced Computing Research for providing the
supercomputer facilities and the CPU time to execute our FFT
codes. This work has been partially supported by the National
Science Foundation under Cooperative Agreement PHY-9210038.

\end{document}